\documentclass[showpacs,preprint,prb,aps]{revtex4}
\usepackage{amssymb}
\usepackage{graphicx}
\pdfoutput=1

\begin{document}

\title{Quantum analysis of a nonlinear microwave cavity-embedded dc SQUID displacement detector}

\author{P. D. Nation}
\author{ M. P. Blencowe}
\affiliation{Department of Physics and Astronomy, Dartmouth College, Hanover, New Hampshire 03755, USA}
\author{E. Buks}
\affiliation{Department of Electrical Engineering, Technion, Haifa 32000 Israel}
\begin{abstract}
We carry out a quantum analysis of a dc SQUID mechanical displacement detector, comprising a SQUID with mechanically compliant loop segment, which is embedded in a microwave transmission line resonator. The SQUID is approximated as a nonlinear, current dependent inductance, inducing an external flux tunable, nonlinear Duffing self-interaction term in the microwave resonator mode equation. Motion of the compliant SQUID loop segment is transduced inductively through changes in the external flux threading SQUID loop, giving a ponderomotive, radiation pressure type coupling between the microwave and mechanical resonator modes.  Expressions are derived for the detector signal response and noise, and it is found that a soft-spring Duffing self-interaction enables a closer approach to the displacement detection standard quantum limit, as well as cooling closer to the ground state.
\end{abstract}


\maketitle

\section{\label{sec:introduction}Introduction}
Recently there has been interest in exploiting the nonlinear dynamics of nanoelectromechanical systems (NEMS) for amplification.\cite{Krommer:2000p2946,Aldridge:2005p3614,Almog:2007p2875}  The use of nonlinear mechanical resonators to some extent parallels investigations with systems comprising purely electronic degrees of freedom, such as  nonlinear superconducting devices incorporating  Josephson Junctions (JJ).\cite{Siddiqi:2004p1662,Lupascu:2006p3289,Lee:2007p3227} For example, it was shown that the bistable response of an RF-driven JJ can be employed as a low noise, high-sensitivity amplifier for superconducting qubits.\cite{Siddiqi:2004p1662}  A similar setup consisting of a JJ embedded in a microwave cavity was used to measure the states of a quantronium qubit,\cite{Metcalfe:2007p1060} where the relevant cavity mode was found to obey the Duffing oscillator equation.\cite{Boaknin:2007p3308}

One area of nanomechanics that has yet to fully explore the possibility of exploiting nonlinearities for sensitive detection involves setups in which a nanomechanical resonator couples either capacitively\cite{Xue:2007p2711,Regal:2008p2560,Teufel:2008p2398} or inductively\cite{Blencowe:2007p285,Buks:2007p279} to a superconducting microwave transmission line resonator, combining elements from both the above-described NEMS and superconducting systems.  Such setups are in some sense the solid-state analogues of optomechanical systems, which ponderomotively couple a movable mirror to the optical field inside a cavity using radiation pressure.\cite{Mancini:1994p2815,LAW:1995p3039,Jacobs:1999p2816,Metzger:2004p2871,Gigan:2006p1942,Arcizet:2006p2873,Schliesser:2006p2558,Mavalvala:2007p160801,Thompson:2008p72}  In both areas, the focus has primarily been on operating in the regime where the cavity and resonator behave to a good approximation as harmonic oscillators interacting via their mutual ponderomotive coupling.  However, in the case of microwave cavities, introducing an embedded JJ,\cite{Boaknin:2007p3308} or simply driving the cavity close to the superconducting critical temperature,\cite{Tholen:2007p1916} results in the breakdown of  the harmonic mode approximation; nonlinear dynamical behavior of the cavity must be taken into account. Furthermore, the ponderomotive coupling term between the microwave or optical cavity mode and mechanical mode is by itself  capable of inducing strong, effective nonlinearities in the respective mode equations.  In optical systems, such nonlinearities can manifest themselves in the appearance of  a bistable (or even multistable)  region for the movable mirror.\cite{DORSEL:1983p2818,Marquardt:2006p103901}  By restricting ourselves to linear microwave cavities, we are overlooking a range of nonlinear phenomena that might enable a closer approach to quantum-limited detection, as well as cooling of the mechanical oscillator closer to its ground state. As an illustration, consider the phase sensitive Josephson parametric amplifier,\cite{yurke:1989p2519,Yurke:2006p1911,bergeal:2008,castellanos:2008}    which exploits the nonlinear effective inductance of the JJ to perform  (in principle) noiseless amplification and  quantum squeezing of the  respective complimentary quadrature amplitudes of the signal oscillator.

In this paper, we will go beyond the usually considered ponderomotively-coupled two oscillator system to include a Duffing nonlinearity in the microwave cavity mode equations. The closed system model Hamiltonian describing the nonlinear microwave-coupled mechanical oscillators is given by Eq.~(\ref{eq:hamiltonian-closed}). The nonlinear microwave mode is externally driven with a pump frequency $\omega_p$ that can be detuned from the transmission line mode frequency $\omega_T$. 
Our investigation will focus on the nonlinear amplifier created by embedding a dc-SQUID displacement detector into a superconducting microwave transmission line.\cite{Blencowe:2007p285}  This has the advantage of significantly larger coupling strengths\cite{Devoret:2007p1782} as compared with existing geometrical coupling schemes.\cite{Xue:2007p2711,Regal:2008p2560,Teufel:2008p2398}  The displacement detector comprises a SQUID with one segment consisting of a doubly-clamped mechanical resonator as shown in Fig.~\ref{fig:diagram}.  The net flux, and therefore circulating current, is modulated by the mechanical motion, providing displacement transduction.  The capacitively-coupled  pump/probe feedline both drives and provides readout of the relevant transmission line resonator mode amplitude (or phase). We will assume transmission line losses are predominantly due  to coupling with the feedline, and that the pump drive is coherent.   The main irreducible noise source is therefore microwave photon shot noise from the drive that acts back on the mechanical oscillator via the intermediate nonlinear microwave resonator and SQUID. Environmental influences on the mechanical oscillator other than that due to the SQUID detector are simply modelled as a free oscillator thermal bath.  By operating the amplifier well below the superconducting critical temperature, and with transmission line currents less than the SQUID JJ's critical current threshold, resistive tunneling of electrons and the associated noise is a negligible contribution.  Similar setups involving JJ elements have been considered previously.\cite{Blencowe:2007p285,Buks:2007p279,Xue:2007p2454,Wang:2007p2819}

With JJ plasma frequencies assumed to be larger than both the mechanical and transmission line fundamental mode frequencies, the SQUID can be considered as a passive, effective inductance element that depends on both the externally applied flux and drive current.  The effective inductance can therefore be freely tuned by varying these external parameters.  Previously, we considered only the lowest, zeroth order expansion of the inductance with respect to the current entering (or exiting) the SQUID,\cite{Blencowe:2007p285} yielding the usual ponderomotively-coupled double harmonic oscillator system.  In this companion paper, we include  the next leading, quadratic order  term, resulting in a nonlinear current dependent inductance. Provided that the current magnitude is small as compared with the JJ's critical current, neglecting higher order terms should not introduce significant errors. The nonlinear inductance induces an effective Duffing (i.e., cubic) self-interaction term in the microwave mode equations of motion.  The results presented here apply  to a broad class of bosonic detector, which includes optomechanical amplifiers with nonlinear cavities\cite{Drummond:1980p3191} that are describable by Hamiltonian~(\ref{eq:hamiltonian-closed}). A related analysis of quantum  noise in a Duffing oscillator amplifier is  given in Ref.~\onlinecite{BabourinaBrooks:2008p3350}.  

The paper is organized as follows.  In Sec.~\ref{sec:motion}  we first derive the truncated Hamiltonian~(\ref{eq:hamiltonian-closed}) that describes the closed system dynamics of the coupled cavity and mechanical resonator fundamental modes. We then derive the quantum Langevin equations of motion that describe the open system dynamics in the presence of the pump/probe line and mechanical oscillator's external environment.  In Sec.~\ref{sec:response} we find expressions for the detector signal response and noise using a semiclassical treatment of the detector's linear response to the external noise input signal driving the mechanical oscillator.  Section~\ref{sec:bistability} determines the critical drive current for the onset of bistability (not to be confused with the JJ critical current).  Sections \ref{sec:detection} and \ref{sec:cooling} discuss the effects of the microwave mode Duffing nonlinearity on mechanical mode displacement detection and cooling, respectively, giving illustrative examples  assuming achievable device parameters.  Section~\ref{sec:conclusion} briefly concludes, while the more technical aspects of the signal and noise term derivations are relegated to the appendix.

\section{\label{sec:motion}Equations of Motion}
\subsection{\label{subsec:closed}Closed System Hamiltonian}

\begin{figure}[htbp]
\begin{center}
\includegraphics[width=5.0in]{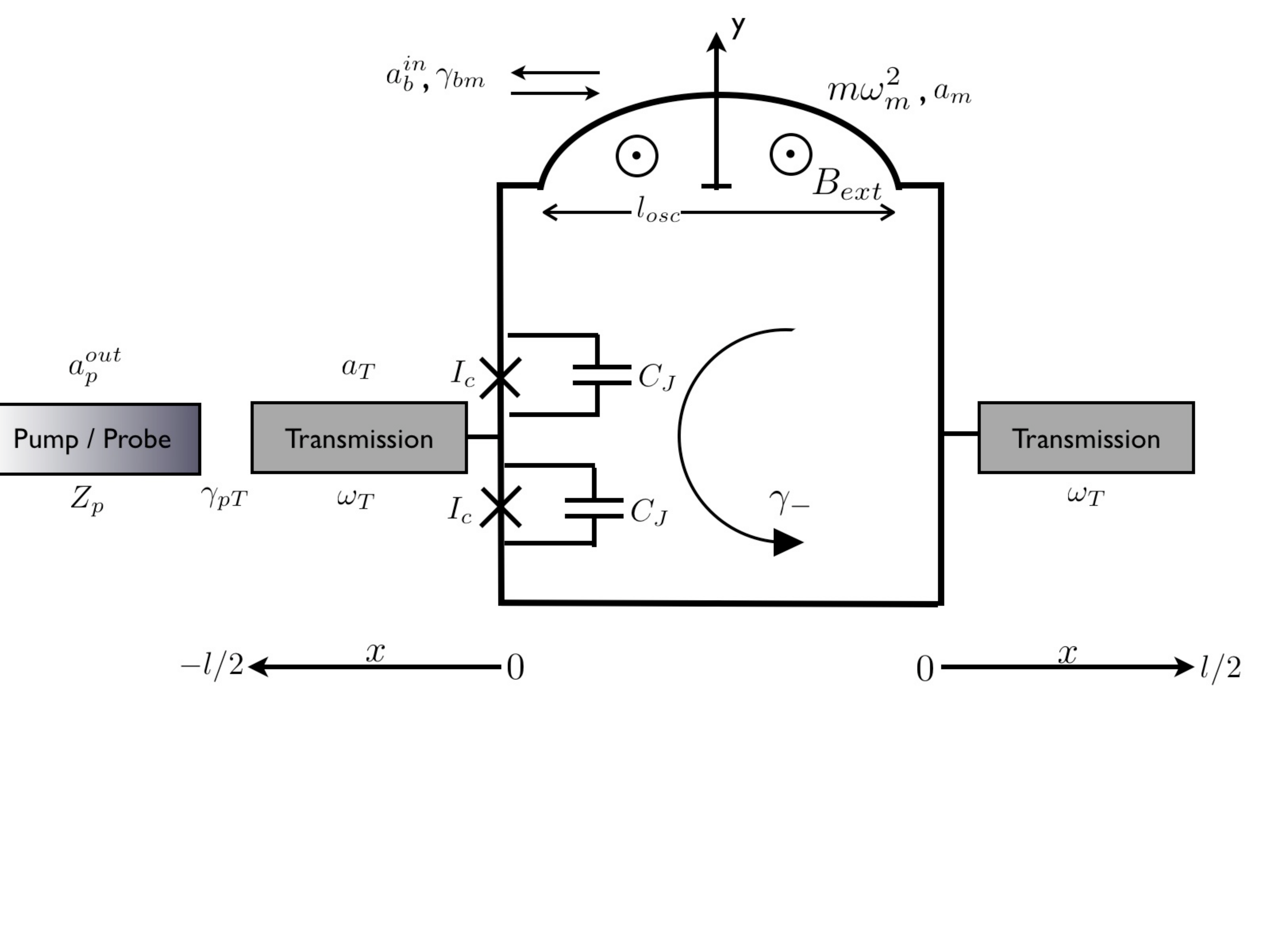}
\caption{Layout of the  dc  SQUID displacement detector.  The dimensions of the SQUID have been enlarged relative to the transmission line to show the key characteristics employed in the analysis.}
\label{fig:diagram}
\end{center}
\end{figure}

The displacement detector scheme is shown in Fig. \ref{fig:diagram}.  The device consists of a stripline resonator (transmission line) of length $l$ bisected by a SQUID.  The transmission line is characterized by an inductance and capacitance per unit length $L_{T}$ and $C_{T}$ respectively.  The SQUID comprises two identical Josephson junctions with critical current $I_{C}$ and capacitance $C_{J}$.  A segment of the SQUID loop is mechanically compliant, forming a doubly clamped resonator of length $l_{\mathrm{osc}}$. We only take into account mechanical fundamental mode displacements in the plane of the loop and assume that the resonator can be modeled effectively as a harmonic oscillator with the $y$ coordinate giving the centre of mass displacement.  The magnetic flux threading the loop is given by $\Phi_{\mathrm{ext}}(y)=\Phi_{\mathrm{ext}}(0)+\lambda B_{\mathrm{ext}}l_{\mathrm{osc}}y$,   where $\Phi_{\mathrm{ext}}(0)\equiv\Phi_{\mathrm{ext}}$ is the flux with the mechanical oscillator fixed at $y=0$, $B_{\mathrm{ext}}$ is the externally applied field in the vicinity of the resonator, and $\lambda$ is a geometrical factor that corrects for the non-uniform displacement of the oscillator along its length.  The coupling between mechanical oscillator and external heat bath is characterized by the oscillator amplitude damping rate $\gamma_{bm}$, while the pump-probe line-transmission line coupling is characterized by the transmission line amplitude damping rate  $\gamma_{pT}$.  In what follows, we will assume weak couplings (i.e., large quality factors for the transmission line and mechanical oscillator) and also that the dominant dissipation mechanism for the transmission line is due to its coupling to the pump-probe line, $\gamma_{pT}$.

In analyzing the SQUID dynamics, an appropriate choice of variables is $\gamma_{\pm}=(\phi_{1}\pm\phi_{2})/2$, where $\phi_{1}$ and $\phi_{2}$ are the gauge invariant phases across the Josephson junctions,\cite{orlando}  while for the transmission line we choose  the phase field $\phi(x,t)$. The transmission line current and voltage are described in terms of $\phi(x,t)$ using the standard telegraphic relations:
\begin{eqnarray}
I(x,t)&=&-\frac{\Phi_{0}}{2\pi L_{T}}\frac{\partial \phi(x,t)}{\partial x}, \\
V(x,t)&=&\frac{\Phi_{0}}{2\pi}\frac{\partial \phi(x,t)}{\partial t},
\label{eq:currentvoltage}
\end{eqnarray}
where $\Phi_{0}=h/(2e)$ is the flux quantum.  Assuming that the SQUID can be lumped at the midpoint $x=0$ of the transmission line, the equations of motion for the closed system comprising the SQUID, transmission line and mechanical oscillator are given by\cite{Buks:2006p346}
\begin{equation}\label{eq:wave}
\frac{\partial^{2}\phi}{\partial t^{2}}=\frac{1}{L_{T}C_{T}}\frac{\partial^{2}\phi}{\partial x^{2}},
\end{equation}
\begin{equation}\label{eq:current-circ}
{\omega_{J}^{-2}}\ddot{\gamma}_{-}+\cos(\gamma_{+})\sin(\gamma_{-})+{2}{\beta^{-1}_{L}}\left[\gamma_{-}-\pi\left(n+\frac{\Phi_{\mathrm{ext}}+\lambda B_{\mathrm{ext}}l_{\mathrm{osc}} y}{\Phi_{0}}\right)\right]=0,
\end{equation}
\begin{equation}\label{eq:current-avg}
{\omega_{J}^{-2}}\ddot{\gamma}_{+}+\sin(\gamma_{+})\cos(\gamma_{-})+\frac{\Phi_{0}}{4\pi L_{T}I_{c}}\frac{\partial \phi(0,t)}{\partial x}=0,
\end{equation}
and
\begin{equation}\label{eq:force}
m\ddot{y}+m\omega_{m}^{2}y-\frac{\Phi_{0}}{\pi L}\lambda B_{\mathrm{ext}}l_{\mathrm{osc}}\gamma_{-}=0,
\end{equation}
where $\omega_{J}=\sqrt{2\pi I_{c}/(C_{J}\Phi_{0})}$ is the plasma frequency of the Josephson junctions, $\beta_{L}\equiv 2\pi L I_{c}/\Phi_{0}$ is a dimensionless parameter with $L$ the SQUID loop self-inductance and $I_c$ the Josephson junction critical current, and where $n$ takes on integer values arising from the requirement that the phase around the loop be single-valued.  Equation (\ref{eq:wave}) is the wave equation for the transmission line,  equation~(\ref{eq:current-circ}) describes the current circling the loop, which depends on the external flux and oscillator position,  equation~(\ref{eq:current-avg}) describes the average current in the loop, and  Eq.~(\ref{eq:force}) is Newton's second law for the mechanical oscillator with  Lorentz force acting on the oscillator.   The  current and voltage across the SQUID must also obey the boundary conditions
\begin{eqnarray}
\frac{\partial \phi(\pm l/2,t)}{\partial x}=0;\ \ \frac{\partial \phi(0^{-},t)}{\partial x}=\frac{\partial \phi(0^{+},t)}{\partial x},\label{eq:currentbc}\\
\dot{\gamma}_{+}-\frac{L}{4 L_{T}}\frac{\partial^2 \phi(0,t)}{\partial t\partial x}=\frac{\partial \phi(0^{-},t)}{\partial t}-\frac{\partial \phi(0^{+},t)}{\partial t}.\label{eq:voltagebc}
\end{eqnarray}

Using Eqs.~(\ref{eq:wave})-(\ref{eq:voltagebc}), we shall now derive approximate equations of motion describing a single mode of the transmission line interacting with the mechanical oscillator, where the form of  the interaction between the two oscillators is governed by the SQUID parameters and boundary conditions.  Assume that the following conditions are satisfied: (a)~$\omega_{J}\gg\omega_{T}\gg\omega_{m}$. (b)~$\beta_{L}\ll 1$. (c)~$|I(0,t)/I_{c}|\ll1$.  (d)~$|\lambda B_{\mathrm{ext}}l_{\mathrm{osc}}y/{\Phi_{0}}|\ll 1$.  Condition~(a) states that the SQUID plasma frequency is much larger than the transmission line mode frequency of interest, $\omega_T$, and consequently we shall ignore the SQUID inertia terms in (\ref{eq:current-circ}) and (\ref{eq:current-avg}). Condition~(b) allows us to neglect the SQUID loop self inductance and, together with (a), eliminate $\gamma_{\pm}$ from the equations by expressing them in terms of the transmission line and oscillator coordinates as series expansions in $\beta_{L}$. Conditions (c) and (d) allow us  to expand the above equations in the transmission line current $I(0,t)\equiv I(t)$ at $x=0$ and in the oscillator displacement $y$. Keeping terms to first order in $y$ and to leading, second order in $I$, Eq.~(\ref{eq:force}) for the mechanical oscillator becomes approximately 
\begin{equation}
m\ddot{y}+m\omega_{m}^{2}y-L_{01}I^2/2 =0.
\label{eq:force-approx}
\end{equation}
The voltage boundary condition (\ref{eq:voltagebc}) can be expressed as
\begin{equation}
\frac{\partial}{\partial t}[L(I,y)I]=\frac{\Phi_{0}}{2\pi}\left[\frac{\partial \phi(0^{-},t)}{\partial t}-\frac{\partial \phi(0^{+},t)}{\partial t}\right]\label{eq:voltage},
\end{equation}
where $L(I,y)$ is the effective inductance, which expanded to second order in $I$ takes the form
\begin{equation}\label{eq:inductance}
 L(I,y)=L_{00}+L_{20}\left({I}/{I_{c}}\right)^{2}+L_{01}y,
\end{equation}
where the $L_{ij}$ coefficients are defined as
\begin{eqnarray}
L_{00}&=&\frac{\Phi_{0}}{4\pi I_{c}}\sec\left({\pi\Phi_{\mathrm{ext}}}/{\Phi_{0}}\right)\label{eq:L00}\\
L_{20}&=&\frac{\Phi_{0}}{96\pi I_{c}}\sec^{3}\left({\pi\Phi_{\mathrm{ext}}}/{\Phi_{0}}\right)\label{eq:L20}\\
L_{01}&=&\frac{\lambda B_{\mathrm{ext}}l_{\mathrm{osc}}}{4I_{c}}\sec\left({\pi\Phi_{\mathrm{ext}}}/{\Phi_{0}}\right)\tan\left({\pi\Phi_{\mathrm{ext}}}/{\Phi_{0}}\right).\label{eq:L01}
\end{eqnarray}
Note that we have neglected the $I^2 y$ term in (\ref{eq:inductance}), restricting ourselves to the leading order coupling only between the transmission line and mechanical oscillator, as already stated. The above equations differ from those of the prequel~\cite{Blencowe:2007p285} through the inclusion of the nonlinear, leading order current-dependent contribution [$L_{20}(I/I_c)^2$] to the effective inductance $L(I,y)$.

The nonlinear voltage boundary condition~(\ref{eq:voltage}) with inductance given by Eq.~(\ref{eq:inductance}) generates frequency tripling harmonics of the transmission line resonator mode. Omitting for the time being the mechanical oscillator degree of freedom, a trial perturbative mode solution to the wave equation~(\ref{eq:wave}) that includes the leading harmonic and solves the current boundary conditions~(\ref{eq:currentbc}) is the following:
\begin{equation}\label{eq:phase-field}
\phi(x,t)=\left\{\begin{array}{ccc}
+A\cos(\omega t +\varphi)\cos\left[k(x-l/2)\right]+a A^3 \cos(3\omega t +3\varphi)\cos\left[3k(x-l/2)\right]; x>0\\
-A\cos(\omega t +\varphi)\cos\left[k(x+l/2)\right]-a A^3 \cos(3\omega t +3\varphi)\cos\left[3k(x+l/2)\right]; x<0,
\end{array}\right.
\end{equation}
where $k=k^{(0)} + k^{(1)}$ and $\omega=|k|/\sqrt{L_T C_T}$. The coefficients $a$, $k^{(0)}$ and $k^{(1)}$ are determined by substituting Eq.~(\ref{eq:phase-field}) into the voltage boundary condition~(\ref{eq:voltage}) and solving perturbatively to order $A^3$, with $k^{(1)}$ scaling as $A^2$. We obtain: $a=-1/48$,
\begin{equation}
({k^{(0)}l}/{2})\tan\left({k^{(0)}l}/{2}\right)=\zeta^{-1}
\label{eq:k0}
\end{equation} 
and
\begin{equation}
k^{(1)} l=-\frac{1}{8}\zeta^3 A^2 (k^{(0)}l)^3 \sin^2 \left({k^{(0)}l}/{2}\right),
\label{eq:k1}
\end{equation}
where
\begin{equation}
\zeta=\frac{L_{00}}{L_Tl}=\frac{\Phi_0}{4\pi L_T l I_c} \sec\left({\pi\Phi_{\mathrm{ext}}}/{\Phi_0}\right).
\label{eq:zeta}
\end{equation}
Considering the transmission line phase field at the location $x=-l/2$,  where the field is pumped and probed (see Fig.~\ref{fig:diagram}), the perturbative solution (\ref{eq:phase-field}) can be obtained from the following single mode equation for $\phi(-l/2,t)\equiv\phi(t)$:
\begin{eqnarray}
\frac{d^2\phi}{dt^2}+\omega_T^2 \phi &+&\frac{1}{12}\omega_T^2 \left(1-18\zeta^3\left[\left({k^{(0)}l}/{2}\right)\sin\left({k^{(0)}l}/{2}\right)\right]^2\right)\phi^3\cr
&&+\frac{1}{12}\left(1-2\zeta^3\left[\left({k^{(0)}l}/{2}\right)\sin\left({k^{(0)}l}/{2}\right)\right]^2\right)\frac{d^2(\phi^3)}{dt^2}=0,
\label{eq:nonlinearmode}
\end{eqnarray}
 where $\omega_T=|k^{(0)}|/\sqrt{L_TC_T}$.
The awkward nonlinear term $\ddot{\phi^3}$ can be eliminated by redefining the phase mode coordinate as $\phi=\psi (1+\Gamma\psi^2)$, provided $|\Gamma|\phi^2\ll 1$, where 
\begin{equation}
\Gamma=-\frac{1}{12}\left(1-2\zeta^3\left[\left({k^{(0)}l}/{2}\right)\sin\left({k^{(0)}l}/{2}\right)\right]^2\right).
\label{eq:Gamma}
\end{equation}
The mode equation~(\ref{eq:nonlinearmode}) in terms of the redefined phase coordinate $\psi$ then becomes
\begin{equation}
\ddot{\psi}+\omega^2_T \psi -\frac{4}{3}\omega_T^2\zeta^3 \left[\left({k^{(0)}l}/{2}\right)\sin\left({k^{(0)}l}/{2}\right)\right]^2 \psi^3 =0.
\label{eq:psimode}
\end{equation}
Thus, embedding a SQUID in a microwave transmission line induces a cubic nonlinearity in the effective single mode equations (under the conditions of small currents as compared with the Josephson junction critical current), resulting in the  familiar (undamped) Duffing oscillator. 

We now restore the mechanical degree of freedom $y(t)$ by assuming that for small and slow displacements [conditions (a) and (d) above], the interaction with $\psi$ can be obtained by expanding $\omega_T$ [through its dependence on $\Phi_{\mathrm{ext}}(y)$] to first order in $y$ in Eq.~(\ref{eq:psimode}) to obtain
\begin{eqnarray}
\ddot{\psi}+\omega^2_T \psi &-&\frac{4}{3}\omega_T^2\zeta^3 \left[\left({k^{(0)}l}/{2}\right)\sin\left({k^{(0)}l}/{2}\right)\right]^2 \psi^3\cr
&&=\frac{\lambda B_{\mathrm{ext}} l_{\mathrm{osc}} y}{\left(\Phi_0/\pi\right)}\frac{\Phi_0}{4\pi L_Tl I_c}\tan\left({\pi\Phi_{\mathrm{ext}}}/{\Phi_0}\right)\sec\left({\pi\Phi_{\mathrm{ext}}}/{\Phi_0}\right)\omega^2_T\psi.
\label{eq:mass-psimode}
\end{eqnarray}  
Equation~(\ref{eq:force-approx}) for the mechanical oscillator, together with Eq.~(\ref{eq:mass-psimode}) for the phase coordinate, follow from the Lagrangian:
\begin{eqnarray}
 &&\mathcal{L}(\psi,y,\dot{\psi},\dot{y})=\frac{1}{2}m\dot{y}^{2}-\frac{1}{2}m\omega_{m}^{2}y^{2}+\frac{1}{2}C_{T}l\left(\frac{\Phi_{0}}{2\pi}\right)^{2}\sin^{2}(k_{0}l/2)\cr
 &&\times\left\{\frac{1}{2}\dot{\psi}^{2}(t)-\frac{1}{2}\left[1-\frac{\lambda B_{\mathrm{ext}}l_{\mathrm{osc}}y}{(\Phi_{0}/\pi)}\frac{\Phi_{0}}{4\pi L_{T}lI_{c}}\tan\left({\pi\Phi_{\mathrm{ext}}}/{\Phi_{0}}\right)\sec\left({\pi\Phi_{\mathrm{ext}}}/{\Phi_{0}}\right)\right]\omega_{T}^{2}\psi^{2}(t)\right.\cr
&&\left.+\frac{1}{3}\omega^2_T\zeta^{3}\left[\left({k^{(0)}l}/{2}\right)\sin\left({k^{(0)}l}/{2}\right)\right]^2\psi^{4}\right\}.
\end{eqnarray}
Introduce the phase momentum coordinate $p_{\psi}=\partial{\mathcal{L}}/\partial{\dot{\psi}}=m_{\psi}\dot{\psi}$ and raising (lowering) operators
\begin{eqnarray}
 \hat{a}_{T}^{\pm}&=&\frac{1}{\sqrt{2m_{\psi}\hbar\omega_{T}}}\left(m_{\psi}\omega_{T}\hat{\psi}\mp i\hat{p}_{\psi}\right)\label{eq:aT}\\
  \hat{a}_{m}^{\pm}&=&\frac{1}{\sqrt{2m\hbar\omega_{m}}}\left(m\omega_{m}\hat{y}\mp i\hat{p}_{y}\right)\label{eq:am}
\end{eqnarray}
satisfying the usual commutation relations, where the effective phase mass is $m_{\psi}=\frac{1}{2}C_{T}l\left(\Phi_{0}/2\pi\right)^{2}\sin^{2}(k_{0}l/2)$. In terms of the raising (lowering) operators, the Hamiltonian operator is
 \begin{equation}\label{eq:hamiltonian-closed}
 H=\hbar\omega_{T}{a}_{T}^{+}{a}_{T}+\frac{1}{12}\hbar\omega_{T}K_{d}({a}_{T}^{+}+{a}_{T})^{4}+\hbar\omega_{m}{a}_{m}^{+}{a}_{m}
+\frac{1}{2}\hbar\omega_{T}K_{Tm}({a}_{T}^{+}+{a}_{T})^2({a}_{m}^{+}+{a}_{m}),
 \end{equation}
 where, for notational convenience, hats on the operators and the minus superscript on the lowering operators will be suppressed from now on. The parameter characterizing the strength of the interaction between the transmission line mode and mechanical oscillator mode is
 \begin{equation}
 K_{Tm}=\frac{\lambda B_{\mathrm{ext}}l_{\mathrm{osc}}\Delta_{zp}}{\left(\Phi_{0}/\pi\right)}\frac{\Phi_{0}}{4\pi L_{T}lI_{c}}\tan\left(\pi\Phi_{\mathrm{ext}}/\Phi_{0}\right)\sec\left(\pi\Phi_{\mathrm{ext}}/\Phi_{0}\right),
\label{eq:ktm}
\end{equation}
where $\Delta_{zp}=\sqrt{\hbar/(2m\omega_m)}$ is the zero-point displacement uncertainty.
 The parameter $K_d$ characterizing the strength of the Duffing nonlinear term takes the form
 \begin{equation}
  K_d = -\left(k^{(0)} l\right)^2 \left(\frac{L_{00}}{L_Tl}\right)^3 \left[\frac{(2e)^2/(2C_Tl)}{\hbar\omega_T}\right],
  \label{eq:Kd}
  \end{equation}
which has been written in such a way as to make clear its various dependencies. In particular, $K_d$ depends essentially on the cube of the ratio of the linear SQUID effective inductance $L_{00}$ to transmission line inductance $L_Tl$, as well as on the ratio of the single Cooper pair charging energy to the microwave mode photon energy of the transmission line. Since the strength and sign of the  linear SQUID inductance depends on the external flux  $\Phi_{\mathrm{ext}}$  [see Eq.~(\ref{eq:L00})], it is possible to vary the strength as well as the sign of the Duffing constant by tuning the external flux either side of  $\Phi_0/2$.  Thus, we can have either spring hardening or spring softening of the transmission line oscillator mode.  Previously this flux tunability was observed in the readout of a persistent current qubit.\cite{Lee:2007p3227} Note, however, that the perturbative approximations that go into deriving the above Hamiltonian~(\ref{eq:hamiltonian-closed}) do not allow too close an approach to the singular half-integer flux quantum point. In particular, the validity of the expansions in $I_T$ and $\beta_L$ properly require the following conditions to hold:
\begin{eqnarray}
\left|\frac{I}{I_{c}}\sec\left({\pi\Phi_{\mathrm{ext}}}/{\Phi_{0}}\right)\right|&\ll&1\label{eq:Iccondition}\\
\left|\beta_{L}\sec\left({\pi\Phi_{\mathrm{ext}}}/{\Phi_{0}}\right)\right|&\ll&1\label{eq:betalcondition}.
\end{eqnarray}

As already noted, Eq.~(\ref{eq:hamiltonian-closed}) without the Duffing nonlinearity coincides with the Hamiltonian commonly used to describe the single mode of an optical cavity interacting with a mechanical mirror via the radiation pressure.  However, we have just seen that embedding a SQUID within a microwave transmission line cavity induces a tunable Duffing self-interaction term as well; it is not so easy to achieve a similar, tunable nonlinearity in the optical cavity counterpart. 

\subsection{\label{sec:open}Open System Dynamics}
Up until now we have considered the transmission line, SQUID and mechanical oscillator  as an isolated system. It is straightforward to couple the transmission line to an external pump-probe feedline and mechanical oscillator to a thermal bath using the `in-out' formalism of Gardiner and Collett.\cite{Gardiner:1985p1483} Assuming weak system-bath couplings justify making the rotating wave approximation (RWA), and furthermore making a Markov approximation for the bath dynamics, the following Langevin equations can be derived for the system mode operators in the Heisenberg picture:
\begin{eqnarray}\label{eq:motionm}	
\frac{d{a}_{m}}{dt}&=&-i\omega_{m}{a}_{m}+\frac{i}{\hbar}\sqrt{\frac{\hbar}{2m\omega_{m}}}F_{\mathrm{ext}}(t)-i\omega_{T}K_{Tm}{a}^{+}_{T}{a}_{T}\cr
&&-\gamma_{bm}{a}_{m}-i\sqrt{2\gamma_{bm}}e^{i\phi_{bm}}{a}_{b}^{\mathrm{in}}(t)
\end{eqnarray}
and
\begin{eqnarray}\label{eq:motiont}
\frac{d{a}_{T}}{dt}&=&-i\omega_{T}{a}_{T}-i\omega_{T}K_{d}{a}_{T}^{+}{a}_{T}{a}_{T}-i\omega_{T}K_{Tm}{a}_{T}({a}^{+}_{m}+{a}_{m})\cr
&&-\gamma_{pT}{a}_{T}-i\sqrt{2\gamma_{pT}}e^{i\phi_{pT}}{a}_{p}^{\mathrm{in}}(t),
\end{eqnarray}
where $\gamma_{bm}$ is the mechanical oscillator amplitude damping rate due to coupling to the bath, $\gamma_{pT}$ is the transmission line mode damping rate due to coupling to the pump-probe line, and we have also assumed that the small Duffing coupling $K_d$ and transmission line-mechanical oscillator coupling $K_{Tm}$ justify applying the RWA to the transmission line mode operator terms.
The `in'  bath  and probe line operators are defined as
\begin{equation}\label{eq:inoperators}
	{a}_{i}^{\mathrm{in}}(t)=\frac{1}{\sqrt{2\pi}}\int d\omega e^{-i\omega(t-t_{0})}{a}_{i}(\omega,t_{0}),
\end{equation}
where $t>t_{0}$, with the states of the pump-probe line and oscillator bath assigned at $t_0$, interpreted as the initial time in the past before the measurement commences. For completeness, we have also included a classical, external time-dependent force $F_{\mathrm{ext}}$(t) acting on the mechanical oscillator, although we shall not address the force detection sensitivity in the present work.

It will be convenient to work with the Fourier transformed Langevin equations. With $O(\omega)=\frac{1}{\sqrt{2\pi}}\int_{-\infty}^{\infty}d\omega e^{i\omega t} O(t)$, Eqs.~(\ref{eq:motionm}) and (\ref{eq:motiont}) become 
\begin{eqnarray}\label{eq:amw}
	{a}_{m}(\omega)&=&\frac{1}{\omega-\omega_{m}+i\gamma_{bm}}\left\{\sqrt{2\gamma_{bm}}e^{i\phi_{bm}}{a}_{b}^{\mathrm{in}}(\omega)-\frac{1}{\sqrt{2m\hbar\omega_{m}}}F_{\mathrm{ext}}(\omega)\right.\cr
	&&\left.+\frac{\omega_{T}K_{Tm}}{2\sqrt{2\pi}}\int_{-\infty}^{\infty}d\omega'
\left[{a}_{T}(\omega'){a}_{T}^{+}(\omega'-\omega)+{a}_{T}^{+}(\omega'){a}_{T}(\omega'+\omega)\right]\right\}
 \end{eqnarray}
 and
 \begin{eqnarray}\label{eq:atw}
{a}_{T}(\omega)&=&\frac{1}{\omega-\omega_{T}+i\gamma_{pT}}\left\{\frac{}{}\sqrt{2\gamma_{pT}}e^{i\phi_{pT}}{a}_{p}^{\mathrm{in}}(\omega)+\frac{\omega_{T}K_{d}}{2\pi}\right.\cr
&&\left.\times \int_{-\infty}^{\infty}d\omega'\int_{-\infty}^{\infty}d\omega''{a}_{T}^{+}(\omega''){a}_{T}(\omega'){a}_{T}(\omega+\omega''-\omega')\right.\cr
&&\left.+\frac{\omega_{T}K_{Tm}}{\sqrt{2\pi}}\int_{-\infty}^{\infty}d\omega'{a}_{T}(\omega') \left[{a}_{m}(\omega-\omega')+{a}_{m}^{+}(\omega'-\omega)\right]\right\}.
\end{eqnarray}

\section{\label{sec:response}Detector Response}
The probe line observables are expressed in terms of the `out' mode operator:
\begin{equation}\label{eq:outoperator}
	{a}_{p}^{\mathrm{out}}(t)=\frac{1}{\sqrt{2\pi}}\int d\omega e^{-i\omega(t-t_1)}{a}_{p}(\omega,t_1),
\end{equation}
where $t_1>t$. The  `out' and `in' probe operators are related via the following useful identity:\cite{Gardiner:1985p1483}
\begin{equation}\label{eq:inoutidentity}
 {a}_{p}^{\mathrm{out}}(t)=-i\sqrt{2\gamma_{pT}}e^{-i\phi_{pT}}{a}_{T}(t)+{a}_{p}^{\mathrm{in}}(t), 
\end{equation}  
which allows us to obtain the expectation value of  a given observable once $a_T(t)$ is determined. 
As illustrative expectation value, we shall consider the variance in the probe line reflected current in a given bandwidth $\delta\omega$ centered about the signal frequency of interest $\omega_s$:\cite{Blencowe:2007p285}
\begin{eqnarray}
\overline{\langle\left[{\delta I}^{\mathrm{out}}(\omega_{s},\delta\omega)\right]^{2}\rangle}&=&\frac{1}{Z_{p}}\int_{\omega_{s}-\delta\omega/2}^{\omega_{s}+\delta\omega/2}\frac{d\omega_{1}d\omega_{2}}{2\pi}\hbar\omega_{1}\left(\frac{2\sin\left[(\omega_{1}-\omega_{2})T_{M}/2\right]}{(\omega_{1}-\omega_{2})T_{M}}\right)\cr
&&\times\frac{1}{2}\langle{a}_{p}^{\mathrm{out}}(\omega_{1}){a}_{p}^{{\mathrm{out}}+}(\omega_{2})+{a}_{p}^{{\mathrm{out}}+}(\omega_{2}){a}_{p}^{\mathrm{out}}(\omega_{1})\rangle,
\label{eq:currentvariance}
\end{eqnarray}
where, in addition to the ensemble average, there is also a time average denoted by the overbar, with the averaging time  taken to be the duration of the measurement  $T_M$, assumed much longer than all other timescales associated with the detector dynamics. In particular, time averaging is required when $F_{\mathrm{ext}}(t)$ has a deterministic time dependence.\cite{Blencowe:2007p285} Expectation values of other observables, such as the reflected voltage variance and reflected power are simply obtained from Eq.~(\ref{eq:currentvariance}) with appropriate inclusions of the probe line impedance $Z_p=\sqrt{L_p/C_p}$: $P^{\mathrm{out}}=\overline{\langle [\delta V^{\mathrm{out}}]^2\rangle}/Z_p=\overline{\langle [\delta I^{\mathrm{out}}]^2\rangle}  Z_p$.  

From the form of the $K_{Tm}$ coupling term in Eq.~(\ref{eq:motiont}), we can see that the motion of the mechanical resonator modulates the transmission line frequency, and thus a complimentary way to transduce displacements besides measuring the current amplitude, is to measure the frequency-dependent, relative phase shift between the `in' pump drive current  and `out' probe current using the homodyne detection procedure.\cite{gardiner} While we shall focus on amplitude detection, the homodyne method can be straighforwardly addressed and is expected to give similar results for the quantum limited detection sensitivity.

Substituting Eq.~(\ref{eq:amw}) into (\ref{eq:atw}), we obtained the following single equation for the transmission line mode operator $a_T$:
\begin{eqnarray}\label{eq:atwonly}
	{a}_{T}(\omega)&=&\int_{-\infty}^{\infty}d\omega'{a}_{T}(\omega-\omega')A(\omega,\omega')+\int_{-\infty}^{\infty}d\omega'B(\omega,\omega'){a}_{T}(\omega-\omega')\cr
	&&\times\int_{-\infty}^{\infty}d\omega''\left[{a}_{T}(\omega''){a}_{T}^{+}(\omega''-\omega')+{a}_{T}^{+}(\omega''){a}_{T}(\omega''+\omega')\right]\cr
&&+D(\omega)\int_{-\infty}^{\infty}d\omega''\int_{-\infty}^{\infty}d\omega'{a}^{+}_{T}(\omega''){a}_{T}(\omega'){a}_{T}(\omega+\omega''-\omega')+C(\omega),
\end{eqnarray}
where
\begin{equation}\label{eq:A}	
	A(\omega,\omega')=\frac{\omega_T K_{Tm}}{\sqrt{2\pi}}\frac{1}{\omega-\omega_{T}+i\gamma_{pT}}
	\left[\frac{S_{m}(\omega')}{\omega'-\omega_{m}+i\gamma_{bm}}+\frac{S_{m}^{+}(-\omega')}{-\omega'-\omega_{m}-i\gamma_{bm}}\right],
\end{equation}
\begin{equation}\label{eq:B}	
	B(\omega,\omega')=\frac{(\omega_T K_{Tm})^{2}}{4\pi}\frac{1}{\omega-\omega_{T}+i\gamma_{pT}} \left[\frac{1}{\omega'-\omega_{m}+i\gamma_{bm}}+\frac{1}{-\omega'-\omega_{m}-i\gamma_{bm}}\right],
\end{equation}
\begin{equation}\label{eq:C}
	C(\omega)=\frac{S_{T}(\omega)}{\omega-\omega_{T}+i\gamma_{pT}}
\end{equation}
and
\begin{equation}\label{eq:D}
	D(\omega)=\frac{\omega_T K_d}{2\pi}\frac{1}{\omega-\omega_{T}+i\gamma_{pT}},
\end{equation}
with mechanical signal operator
\begin{equation}
	S_{m}(\omega)=\sqrt{2\gamma_{bm}}e^{i\phi_{bm}}{a}_{b}^{\mathrm{in}}(\omega)-\frac{1}{\sqrt{2m\hbar\omega_{m}}}F_{\mathrm{ext}}(\omega)\label{eq:sm}
\end{equation}
and noise operator
\begin{equation}
	S_{T}(\omega)=\sqrt{2\gamma_{pT}}e^{i\phi_{pT}}{a}_{p}^{\mathrm{in}}(\omega).\label{eq:st}
\end{equation}
For small signal strength, it is assumed that Eq.~(\ref{eq:atwonly}) can be solved as a series expansion up to first order in $A(\omega,\omega')$, giving the usual linear-response approximation. I.e., $a_T(\omega)\approx a_T^{(0)}(\omega)+a_T^{(1)}(\omega)$, where the noise component  $a_T^{(0)}(\omega)$ is the solution to Eq.~(\ref{eq:atwonly}) with the mechanical signal source term $A(\omega,\omega')$ set to zero, while the signal component $a_T^{(1)}(\omega)$ is the part of the solution to Eq.~(\ref{eq:atwonly}) that depends linearly on $A(\omega,\omega')$. Thus, from Eq.~(\ref{eq:inoutidentity}) we can express the `out' probe mode operator as follows:
\begin{equation}
a_p^{\mathrm{out}}(\omega)=\left[-i\sqrt{2\gamma_{pT}}e^{-i\phi_{pT}} a^{(1)}_T (\omega)\right]+\left[-i\sqrt{2\gamma_{pT}}e^{-i\phi_{pT}} a^{(0)}_T (\omega)+a_p^{\mathrm{in}}(\omega)\right],
\label{eq:asignalnoise}
\end{equation}
where the first square bracketed term gives the signal contribution to the detector response and the second square bracketed term gives the noise contribution. 

As `in' states, we consider the mechanical oscillator bath to be in a thermal state at temperature $T$ and the pump line to be in a coherent state centered about the pump frequency $\omega_p$:\cite{Johansson:2006p2429}
\begin{equation}\label{eq:coherent}
\left|\{\alpha(\omega)\}\right.\rangle_{p}=\exp\left(\int d\omega\alpha(\omega)\left[{a}_{p}^{\mathrm{in}+}(\omega)-{a}_{p}^{\mathrm{in}}(\omega)\right]\right)\left|0\right.\rangle_{p},
\end{equation}
where $\left|0\right.\rangle_{p}$ is the vacuum state and 
\begin{equation}\label{eq:alpha}
\alpha(\omega)=-I_{0}\sqrt{\frac{Z_{p}T^{2}_{M}}{2\hbar}}\frac{e^{-(\omega-\omega_{p})^{2}T^{2}_{M}/2}}{\sqrt{\omega}}.
\end{equation}
The coherent state coordinate $\alpha(\omega)$ is parametrized such that the expectation value of the right-propagating `in' current $I^{\mathrm{in}}(x,t)$ with respect to this coherent state has amplitude $I_0$, where
\begin{equation}
I^{\mathrm{in}}(x,t)=i\sqrt{\frac{\hbar}{4\pi Z_p}}\int_0^{\infty} d\omega \sqrt{\omega} \left[e^{i\omega (x/v_p -t)} a_p^{\mathrm{in}}(\omega)-e^{-i\omega (x/v_p -t)} a_p^{\mathrm{in}+}(\omega)\right],
\label{eq:incurrentoperator}
\end{equation}
with $v_p=1/\sqrt{L_pC_p}$ the wave propagation velocity in the pump probe line.

With the pump probe line in a coherent state, we assume that for large drive currents  Eq.~(\ref{eq:atwonly}) can be approximately solved  using a semiclassical, `mean field' approximation, where the quantum fluctuation $\delta a_T^{(0)}(\omega)$ in $a_T^{(0)}(\omega)=\langle a_T^{(0)}(\omega)\rangle +\delta a_T^{(0)}(\omega)$ is kept to first order only. However,  the nonlinear Duffing  and transmission line-mechanical oscillator interaction terms can give rise to a bistability in the transmission line oscillator dynamics and one must be careful when interpreting the results from the mean field approximation when operating close to a bifurcation point; large fluctuations can occur in the oscillator amplitude as it jumps between the two metastable amplitudes, which are not accounted for in the mean field approximation. (See Refs. \onlinecite{Dykman:1980p480} and~\onlinecite{Dykman:2007p1864} for respective analyses of the classical and quantum oscillator fluctuation dynamics near a bifurcation point).  This issue will be further discussed in the following sections. 

The solutions to the signal $a_T^{(1)}(\omega)$ and noise $a_T^{(0)}$ terms parallel closely our previous calculations, which omitted the Duffing nonlinearity;\cite{Blencowe:2007p285} the Duffing ($D$) term in Eq.~(\ref{eq:atwonly}) has a very similar form to the transmission line oscillator coupling ($B$) term, both involving $a_T^2 a_T^+$ operator combinations. We therefore relegate the solution details to the appendix, presenting only the essential results in this section.

The solution to $\langle a _T^{(0)}(\omega)\rangle$ is sharply peaked about the pump frequency $\omega_p$ for large $T_M$   and so can be approximately expressed as a delta function:  $\langle a _T^{(0)}(\omega)\rangle=\chi \delta (\omega-\omega_p)$. Substituting this expression into Eq.~(\ref{eq:zeroth-atw-coherent}), we obtain for the  amplitude $\chi$: 
\begin{equation}\label{eq:mean-field0}
	\chi=c+\left[2B(\omega_{p},0)+D(\omega_{p})\right]\chi\left|\chi\right|^{2},
\end{equation}
with
\begin{equation}
	c=\frac{i\sqrt{2\pi}e^{i\phi_{pT}}}{\gamma_{pT}-i\Delta\omega}\sqrt{\frac{I_{0}^{2}Z_{p}\gamma_{pT}}{\hbar\omega_{p}}}, \label{eq:c}
\end{equation}
where $\Delta\omega=\omega_{p}-\omega_{T}$ is the detuning of the pump frequency $\omega_p$ from the transmission line resonance frequency $\omega_T$. Using the expressions for $B(\omega_p,0)$ and $D(\omega_p)$, Eq.~(\ref{eq:mean-field0}) can be written as
\begin{equation}
(\omega_T-\omega_p -i \gamma_{pT})\chi+\frac{\omega_T}{2\pi} {\mathcal{K}}\chi |\chi|^2 = e^{i\phi_{pT}}\sqrt{{{2\pi}I_0^2 Z_p \gamma_{pT}}/{(\hbar\omega_p})},\label{eq:meanfield} 
\end{equation}
where the effective Duffing coupling is defined as
\begin{equation}
{\mathcal{K}}=K_d -\frac{2 \omega_T \omega_m}{\omega_m^2 +\gamma_{bm}^2} K^2_{Tm}.
\label{eq:k}
\end{equation}
Notice that the interaction between the transmission line and mechanical oscillator induces an additional Duffing nonlinearity (the second term involving $K_{Tm}$ in ${\mathcal{K}}$) in the transmission line mode amplitude effective equations of motion (\ref{eq:meanfield}).  However, in contrast with the $K_d$ nonlinearity, which can be tuned to have either sign, the former mechanically-induced nonlinearity is always negative and thus has a ``spring-softening" affect on the transmission line mode. Interestingly, by choosing an appropriate compensating ``spring hardening" $K_d>0$, the effective Duffing constant ${\mathcal{K}}$ can in principle be completely suppressed so that the next non-vanishing higher order nonlinearity would govern the mode amplitude dynamics.

Once we have the solution for $\langle a _T^{(0)}(\omega)\rangle$, the solutions for the quantum signal $a_T^{(1)}(\omega)$ and quantum noise $\delta a_T^{(0)}(\omega)$ are obtained from Eqs.~(\ref{eq:first-atw}) and (\ref{eq:zeroth-atw-quantum}), respectively. These solutions can be expressed as follows:
\begin{equation}
{a}_{T}^{(1)}(\omega)=\alpha_{1}(\omega)A(\omega,\omega-\omega_{p})+\alpha_{2}(\omega)A(\omega-2\Delta\omega,\omega-\omega_{p})\label{eq:a1solution}
\end{equation}
and
\begin{equation}\label{eq:a0solution}
\delta{a}_{T}^{(0)}(\omega)=\beta_{1}(\omega)\delta C(\omega)+\beta_{2}(\omega)\delta C^{+}(2\omega_{p}-\omega),
\end{equation}
where the $\alpha_{i}(\omega)$ and $\beta_{i}(\omega)$ functions are defined in Eqs.~(\ref{eq:alpha1}), (\ref{eq:alpha2}), (\ref{eq:beta1}), and (\ref{eq:beta2}).

Substituting Eqs.~(\ref{eq:a1solution}) and (\ref{eq:a0solution}) into the expression (\ref{eq:asignalnoise}) for $a^{\mathrm{out}}(\omega)$ and then evaluating the signal component of the detector response~(\ref{eq:currentvariance}), we obtain\cite{Blencowe:2007p285}
\begin{eqnarray}\label{eq:current-signal}
&&\left.\overline{\langle\left[{\delta I}^{\mathrm{out}}(\omega_{s},\delta\omega)\right]^{2}\rangle}\right|_{\mathrm{signal}}=\left(\frac{I_{0}K_{Tm}\omega_{T}}{\gamma_{pT}}\right)^{2}\frac{\gamma_{pT}^{2}}{\gamma_{pT}^{2}+\Delta\omega^{2}}\cr
&&\times\int_{\omega_{s}-\delta\omega/2}^{\omega_{s}+\delta\omega/2}\frac{d\omega}{2\pi}\left(\frac{\omega}{\omega_{p}}\frac{\gamma_{pT}^{2}}{(\omega-\omega_{p}+\Delta\omega)^{2}+\gamma_{pT}^{2}}\right)\left|\frac{\alpha_{1}(\omega)}{c}+\frac{\alpha_{2}(\omega)}{c}\left(\frac{\omega-\omega_{p}+\Delta\omega+i\gamma_{pT}}{\omega-\omega_{p}-\Delta\omega+i\gamma_{pT}}\right)\right|^{2}\cr
&&\times\left(\frac{2\gamma_{bm}}{(\omega-\omega_{p}-\omega_{m})^{2}+\gamma_{bm}^{2}}[2n(\omega-\omega_{p})+1]+\frac{2\gamma_{bm}}{(\omega_{p}-\omega-\omega_{m})^{2}+\gamma_{bm}^{2}}[2n(\omega_{p}-\omega)+1]\right)\cr
&&+\left(\frac{I_{0}K_{Tm}\omega_{T}}{\gamma_{pT}}\right)^{2}\frac{\gamma_{pT}^{2}}{\gamma_{pT}^{2}+\Delta\omega^{2}}\frac{1}{2m\hbar\omega_{m}\gamma_{bm}}\int_{\omega_{s}-\delta\omega/2}^{\omega_{s}+\delta\omega/2}\frac{d\omega d\omega'}{2\pi}\left(\frac{\omega}{\omega_{p}}\frac{\gamma_{pT}^{2}}{(\omega-\omega_{p}+\Delta\omega)^{2}+\gamma_{pT}^{2}}\right)\cr
&&\times\left|\frac{\alpha_{1}(\omega)}{c}+\frac{\alpha_{2}(\omega)}{c}\left(\frac{\omega-\omega_{p}+\Delta\omega+i\gamma_{pT}}{\omega-\omega_{p}-\Delta\omega+i\gamma_{pT}}\right)\right|^{2}\frac{\sin\left[(\omega-\omega')T_{m}/2\right]}{(\omega-\omega')T_{m}/2}\cr
&&\times\left(\frac{2\gamma_{bm}}{(\omega-\omega_{p}+\Delta\omega)^{2}+\gamma_{bm}^{2}}F_{\mathrm{ext}}(\omega-\omega_{p})F^{*}_{\mathrm{ext}}(\omega'-\omega_{p})\right.\cr
&&\left.+\frac{2\gamma_{bm}}{(\omega_{p}-\omega-\omega_{m})^{2}+\gamma_{bm}^{2}}F_{\mathrm{ext}}(\omega_{p}-\omega)F^{*}_{\mathrm{ext}}(\omega_{p}-\omega')\right),	
\end{eqnarray}	
where $n(\omega)=(e^{\hbar\omega/k_BT} -1)^{-1}$ is the thermal average occupation number for bath mode $\omega$. In the limit of small drive current amplitude $I_0\rightarrow 0$, we have $\alpha_1 (\omega)/c \rightarrow 1$, $\alpha_2 (\omega)/c \rightarrow 0$, and we see that the signal spectrum comprises two Lorentzian peaks centered at $\omega_p\pm\omega_m$. The $\omega_p+\omega_m$ peak corresponds to  phase preserving detection, in the sense that $a_p^{\mathrm{out}}$ gives the amplified $a_b^{\mathrm{in}}$ signal, while the  $\omega_p-\omega_m$ peak corresponds to phase conjugating detection, with $a_p^{\mathrm{out}}$ amplifying the $a^{\mathrm{in}+}_b$ signal.\cite{Caves:1982p1311}    Increasing the drive current amplitude causes the peaks to shift, and the peak widths relative to their height to change, signifying renormalization of the mechanical oscillator frequency and damping rate. The noise component of the detector response is
\begin{eqnarray}\label{eq:current-noise}
&&\left.\overline{\langle\left[{\delta I}^{\mathrm{out}}(\omega_{s},\delta\omega)\right]^{2}\rangle}\right|_{\mathrm{noise}}=\frac{1}{Z_{p}}\int_{\omega_{s}-\delta\omega/2}^{\omega_{s}+\delta\omega/2}\frac{d\omega}{2\pi}\hbar\omega\frac{2\gamma_{pT}^{2}}{(\omega-\omega_{p}+\Delta\omega)^{2}+\gamma_{pT}^{2}}\cr
&&\times\left(\left|\beta_{1}(\omega)\right|^{2}+\frac{(\omega-\omega_{p}+\Delta\omega)^{2}+\gamma_{pT}^{2}}{(\omega-\omega_{p}-\Delta\omega)^{2}+\gamma_{pT}^{2}}\left|\beta_{2}(\omega)\right|^{2}
-{\mathrm{Re}}\left[\beta_{1}(\omega)\right]+\frac{(\omega-\omega_{p}+\Delta\omega)}{\gamma_{pT}}{\mathrm{Im}}\left[\beta_{1}(\omega)\right]\right)\cr
&&+Z_{p}^{-1}\frac{\hbar\omega_{s}}{2}\frac{\delta\omega}{2\pi},
\end{eqnarray}
where the integral term involving the $\beta_i (\omega)$ functions includes the back reaction noise  on the mechanical oscillator and the term involving $Z_p$ describes the probe line zero-point fluctuations added at the output.

In Sec.~\ref{sec:detection} we will numerically evaluate Eqs.~(\ref{eq:current-signal}) and (\ref{eq:current-noise}) and in particular compare the detector noise with the minimum noise bound discussed by Caves\cite{{Blencowe:2007p285},{Caves:1982p1311}}  that follows from the Heisenberg uncertainty principle for the detector:
\begin{eqnarray}\label{eq:current-min}
&&\left.\overline{\langle\left[\delta I^{\mathrm{out}}(\omega_{s},\delta\omega)\right]^{2}\rangle}\right|_{\mathrm{min-noise}}=\left|Z_p^{-1}\frac{\hbar\omega_{s}}{2}\frac{\delta\omega}{2\pi}-\left(\frac{I_{0}K_{Tm}\omega_{T}}{\gamma_{pT}}\right)^{2}\frac{\gamma^{2}_{pT}}{\gamma^{2}_{pT}+\Delta\omega^{2}}\right.\cr
&&\left.\times\int_{\omega_{s}-\delta\omega/2}^{\omega_{s}+\delta\omega/2}\frac{d\omega}{2\pi}\left(\frac{\omega}{\omega_{p}}\frac{\gamma_{pT}^{2}}{(\omega-\omega_{p}+\Delta\omega)^{2}+\gamma_{pT}^{2}}\right)\left|\frac{\alpha_{1}(\omega)}{c}+\frac{\alpha_{2}(\omega)}{c}\left(\frac{\omega-\omega_{p}+\Delta\omega+i\gamma_{pT}}{\omega-\omega_{p}-\Delta\omega+i\gamma_{pT}}\right)\right|^{2}\right.\cr
&&\left.\times
\left(\frac{2\gamma_{bm}}{(\omega-\omega_{p}-\omega_{m})^{2}+\gamma^{2}_{bm}}-\frac{2\gamma_{bm}}{(\omega_{p}-\omega-\omega_{m})^{2}+\gamma^{2}_{bm}}\right)\right|.
\end{eqnarray}

\section{\label{sec:bistability}Bistability Conditions}

We have seen [Hamiltonian  (\ref{eq:hamiltonian-closed})]  that the current-dependent SQUID effective inductance gives rise to a transmission line Duffing type nonlinearity with strength $K_d$. Furthermore, there is a nonlinear coupling with strength $K_{Tm}$ between the transmission line and mechanical oscillator.  These two nonlinearities correspond respectively to the cubic terms  proportional  to $K_{d}$ and $K^2_{Tm}$ in the mean transmission line coordinate amplitude  $\chi$ equation (\ref{eq:mean-field0}). For sufficiently large drive current amplitude $I_0$ and/or coupling strengths $K_{Tm}$, $K_d$,  the cubic term $\chi\left|\chi\right|^2$ term in Eq.~(\ref{eq:meanfield})  becomes appreciable, resulting in three real solutions over a certain pump frequency range $\omega_p$.  This parameter regime defines the bistable region of the detector phase space (the intermediate amplitude solution is unstable and cannot be realized in practice).  In the following, we determine the conditions on the parameters for the bistable region  employing the analysis of Ref.~\onlinecite{Yurke:2006p1911}.

\begin{figure}[htbp]
\begin{center}
\includegraphics[width=2.7in]{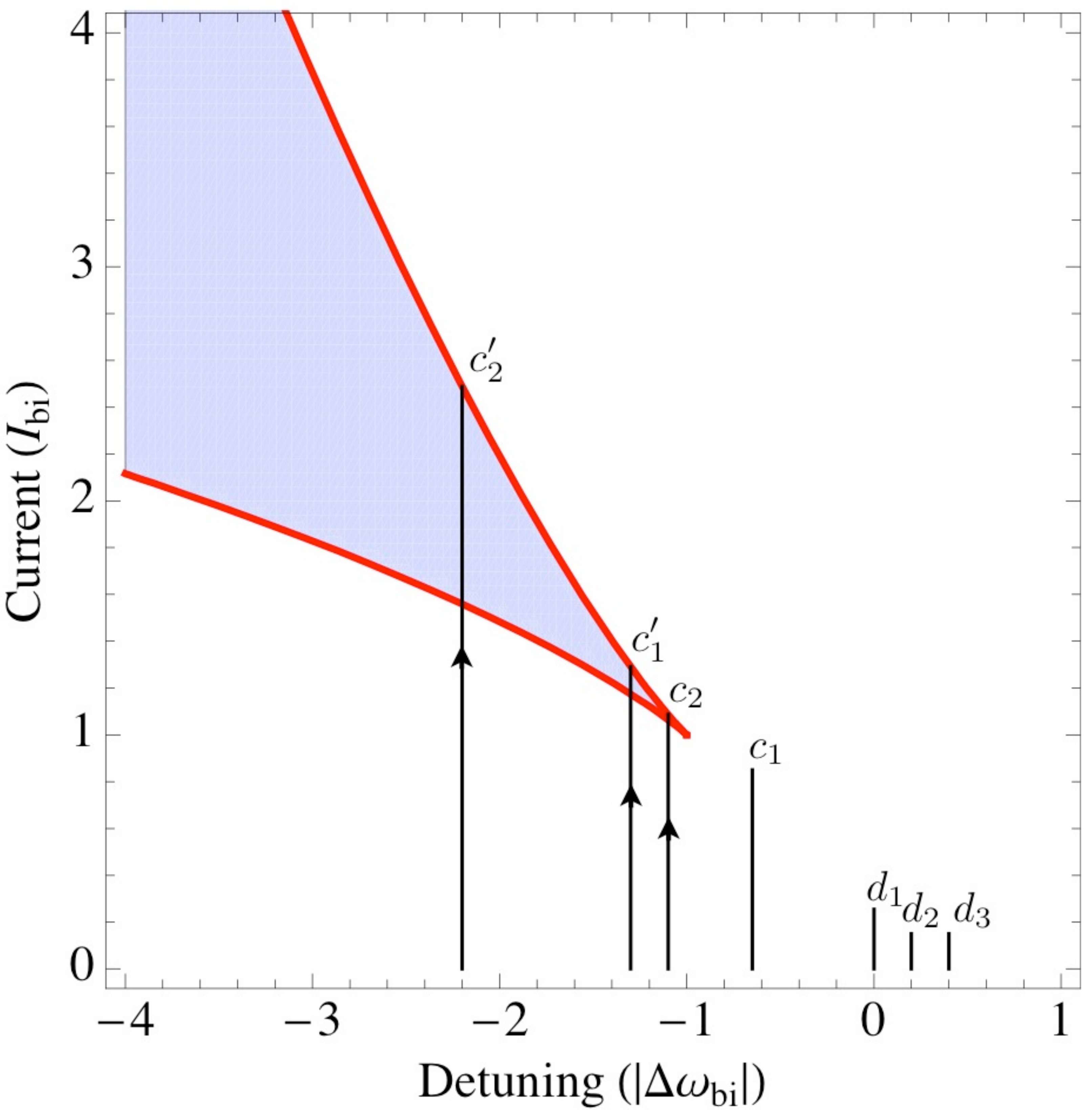}
\caption{Bistable region (shaded) of the cavity-oscillator system for negative, spring softening Duffing nonlinearity.  The drive current and detuning are expressed in units of the bistability onset critical values $I_{bi}$ and $|\Delta\omega_{bi}|$. The labelled straight line traces correspond to  detection ($d$) and cooling ($c$) current drive-detuning parameter examples considered in Secs~\ref{sec:detection} and \ref{sec:cooling}. The arrows give the direction in which the drive current is varied in order to enter the bistable region on the small amplitude branch.}
\label{fig:region}
\end{center}
\end{figure}

We first express the transmission line mode coordinate in terms of its phase and amplitude:
\begin{eqnarray}\label{eq:chi_B}
 \langle a_T^{(0)}(t)\rangle&=&Me^{-i(\omega_{p}t+\phi_{M})},\cr
 \chi&=&\sqrt{2\pi}M e^{-i\phi_M},
 \end{eqnarray}
where the amplitude $M$ is a positive real constant and recall $\chi$ is defined through the relation $ \langle a_T^{(0)}(\omega)\rangle=\chi \delta(\omega-\omega_p)$. Equation~(\ref{eq:meanfield}) then becomes
\begin{equation}\label{eq:kerr-eq}
\left(\omega_{T}-\omega_{p}-i\gamma_{pT}\right)M+\mathcal{K}\omega_T M^{3}=\sqrt{2\gamma_{pT}}\langle b_{pT}^{\mathrm{in}}\rangle e^{i(\phi_{pT}+\phi_{M})},
\end{equation}
where
\begin{equation}\label{eq:bin}
\langle b_{pT}^{\mathrm{in}}\rangle=\sqrt{\frac{I_{0}^{2}Z_{p}}{2\hbar\omega_{p}}}.
\end{equation}
Multiplying both sides of Eq.~(\ref{eq:kerr-eq}) by their complex conjugates and substituting $E=M^{2}$, we obtain  the following third-order polynomial in $E$:
\begin{equation}\label{eq:third_poly}
	E^{3}+\frac{2(\omega_{T}-\omega_{p})}{\omega_T\mathcal{K}}E^{2}+\frac{(\omega_{T}-\omega_{p})^{2}+\gamma_{pT}^{2}}{\omega_T^2\mathcal{K}^{2}}E=\frac{2\gamma_{pT}\langle b_{pT}^{\mathrm{in}}\rangle^{2}}{\omega_T^2\mathcal{K}^{2}}.
\end{equation}
The bifurcation line in current drive and detuning parameter space that delineates between the single solution and bistable solution regions occurs where the susceptibility $\partial E/\partial\omega_{p}$ diverges. If we further impose the condition that the transition between the two regions is continuous, i.e. $\partial^{2}\omega_{p}/\partial^{2}E=0$, we obtain the bistability onset critical point.  From Eq.~(\ref{eq:third_poly}), these two requirements can be written
\begin{eqnarray}
&3\mathcal{K}^{2}E^{2}+4(\omega_{T}-\omega_{p})\mathcal{K}E+(\omega_{T}-\omega_{p})^{2}+\gamma_{pT}^{2}=0,\label{eq:susceptibility} \\
&6\mathcal{K}^{2}E+4(\omega_{T}-\omega_{p})\mathcal{K}=0.
\end{eqnarray}
Solving these equations simultaneously for $E$ and $\Delta\omega=\omega_p-\omega_T$ yields the following bistability onset critical values:
\begin{eqnarray}
E_{{bi}}&=&\frac{2\gamma_{pT}}{\sqrt{3}\omega_T|\mathcal{K}|}\label{eq:bi}, \cr
\Delta\omega_{{bi}}&=&\sqrt{3}\gamma_{pT}\frac{\mathcal{K}}{|\mathcal{K}|}.\label{eq:detune}
\end{eqnarray}
Substituting these critical values into Eq.~(\ref{eq:third_poly}) gives 
\begin{equation}
\langle b_{pT}^{\mathrm{in}}\rangle_{{bi}}^{2}=\frac{4\gamma_{pT}^{2}}{3\sqrt{3}\omega_T|\mathcal{K}|}.
\end{equation}
Finally, using Eq.~(\ref{eq:bin}) we obtain the driving current critical amplitude:
\begin{equation}\label{eq:Ibi}
I_{{bi}}=2\gamma_{pT}\sqrt{\frac{2\hbar\omega_{p}}{3\sqrt{3}\omega_T|\mathcal{K}|Z_{p}}}.
\end{equation}
Note, the requirement that we operate below the Josephson critical current, $I_{0}<I_{c}$, gives a lower limit on the value of $|\mathcal{K}|$ for which our system can approach the bistability onset. 
The boundary of the bistable region that is given by the vanishing susceptibility equation (\ref{eq:susceptibility}) can be expressed in units of the bistability onset critical current $I_{bi}$ and detuning value $\Delta\omega_{bi}$  using Eqs.~(\ref{eq:Ibi}) and~(\ref{eq:detune}) to obtain\cite{Dykman:1980p480}
\begin{equation}\label{eq:boundary}
\frac{I}{I_{bi}}=\frac{1}{2}\left(\frac{\Delta\omega}{\Delta\omega_{bi}}\right)^{3/2}\left\{1+3\left(\frac{\Delta\omega_{bi}}{\Delta\omega}\right)^{2}\pm\left[1-\left(\frac{\Delta\omega_{bi}}{\Delta\omega}\right)^{2}\right]^{3/2}\right\}^{1/2},
\end{equation}
where the $\pm$ roots give the upper and lower boundaries of the bistable region, respectively (see Fig.~\ref{fig:region}).  As mentioned in the preceding section, care must be taken when applying our semiclassical, mean field approximations to the detector signal and noise response when approaching closely the  bifurcation boundary lines. Fluctuation-induced jumps between the small and large amplitude solutions of the transmission line  mode can occur that are not accounted for in the mean field approximation. Nevertheless, in the next two sections we shall in some instances evaluate the detector response close the boundaries of the bistability region. For example, we shall see that significant improvements in cooling can be achieved provided a way is found to keep the transmission line mode on the low amplitude solution branch when operating in the bistable region.  

\section{\label{sec:detection}Displacement Detection}

Assuming that $\gamma_{bm}\ll\gamma_{pT}$, i.e., the unrenormalized mechanical oscillator amplitude damping rate is much smaller  than the transmission line oscillator amplitude damping rate, then the detector spectral noise and response in the mechanical signal bandwidth is approximately white over a large range of drive current and detuning parameter space. The mechanical signal and  noise response spectra are therefore approximately Lorentzian and Eqs.~(\ref{eq:current-signal}) and (\ref{eq:current-noise}) can be parametrized as
\begin{eqnarray}
&&\left.\overline{\langle\left[{\delta I}^{\mathrm{out}}(\omega_{s}=\omega_p\pm R_{\omega}\omega_m,\delta\omega)\right]^{2}\rangle}\right|_{\mathrm{signal}} Z_p\cr
&&=G_{\pm} \frac{\hbar}{2m R_{\omega}\omega_m}\int_{\omega_{s}-\delta\omega/2}^{\omega_{s}+\delta\omega/2}\frac{d\omega}{2\pi} \frac{2\gamma_{bm}[2n(R_{\omega}\omega_m)+1]}{(\omega-\omega_p\mp R_{\omega}\omega_m)^2 +(R_{\gamma} \gamma_{bm})^2}
\label{eq:paramcurrentsignal}
\end{eqnarray} 
and
\begin{eqnarray}
&&\left.\overline{\langle\left[{\delta I}^{\mathrm{out}}(\omega_{s}=\omega_p\pm R_{\omega}\omega_m,\delta\omega)\right]^{2}\rangle}\right|_{\mathrm{noise}} Z_p\cr
&&=G_{\pm} \frac{\hbar}{2m R_{\omega}\omega_m}\int_{\omega_{s}-\delta\omega/2}^{\omega_{s}+\delta\omega/2}\frac{d\omega}{2\pi} \frac{2\gamma_{\mathrm{back}}[2n^{\pm}_{\mathrm{back}}+1]}{(\omega-\omega_p\mp R_{\omega}\omega_m)^2 +(R_{\gamma} \gamma_{bm})^2}\cr
&&+\left.\overline{\langle\left[{\delta I}^{\mathrm{out}}(\omega_{s}=\omega_p\pm R_{\omega}\omega_m,\delta\omega)\right]^{2}\rangle}\right|_{\mathrm{added}~\mathrm{noise}} Z_p,
\label{eq:paramcurrentnoise}
\end{eqnarray} 
where $G_{\pm}$ is the phase preserving (conjugating) gain (in W$\cdot$m$^{-2}$), $n(R_{\omega}\omega_m)$ is the mechanical oscillator's external bath occupation number at its renormalized frequency $R_{\omega} \omega_m$,  $R_{\gamma} \gamma_{bm}$ is the renormalized (i.e., net) mechanical oscillator  damping rate, and the detector back reaction noise on the oscillator is effectively that of a thermal bath  with damping rate $\gamma_{\mathrm{back}}=\gamma_{bm} (R_{\gamma}-1)$ and thermal average occupation number $n^{\pm}_{\mathrm{back}}$. Note, from here on we do not consider an external classical force driving the mechanical oscillator; the focus is on displacement detection rather than force detection. The added noise term in Eq.~(\ref{eq:paramcurrentnoise}) comprises output noise that is not due to the action  of the detector on the mechanical oscillator; the added noise is present even when there is no coupling to the mechanical oscillator, i.e., when $K_{Tm}=0$.  In the absence of the  transmission line Duffing nonlinearity, the added noise simply consists of the probe line zero-point fluctuations $\hbar\omega_s\delta\omega/(4\pi Z_p)$. However, with the Duffing nonlinearity present, the added noise will be in excess of the probe line zero-point fluctuations.

The convenient  Lorentzian parametrization approximations of the mechanical signal~(\ref{eq:paramcurrentsignal}) and noise response spectra~(\ref{eq:paramcurrentnoise}) that provide the above-described  effective thermal description of the back reaction noise will break down as one approaches arbitrarily closely the jump points at the ends of the small or large amplitude transmission line oscillator solution branches occuring at the boundaries of the bistable region indicated in Fig.~\ref{fig:region}. This is a consequence of the diverging damping (i.e., ring-down) time of transmission line mode.\cite{Yurke:2006p1911} Thus, when numerically solving ~(\ref{eq:current-signal}) and (\ref{eq:current-noise}) to extract the effective thermal properties of the detector back reaction, it is important to always check the accuracy of the Lorentzian spectrum approximation.

For sufficiently large gain (i.e., large current drive amplitude), we can neglect the added noise contribution and we have for the noise-to-signal response ratio when the mechanical oscillator external bath is at absolute zero [i.e., $n(R_{\omega} \omega_m)=0$]:
\begin{equation} 
\frac{\langle\left[{\delta I}^{\mathrm{out}}\right]^{2}\rangle_{\mathrm{noise}}}{\langle\left[{\delta I}^{\mathrm{out}}\right]^{2}\rangle_{\mathrm{signal}}}=(2n^{\pm}_{\mathrm{back}}+1)\frac{ \gamma_{\mathrm{back}}}{\gamma_{bm}}.
\label{eq:noisetosignal}
\end{equation}
On the other hand,  in the large gain limit the Caves noise lower bound (\ref{eq:current-min}) gives a noise-to-signal ratio of one. For large gain, we typically have $|2n^{\pm}_{\mathrm{back}}+1|\gg 1$ and thus to approach the Caves bound necessarily requires $|\gamma_{\mathrm{back}}|\ll \gamma_{bm}$.\cite{Clerk:2004p245306}

As an example, we numerically solve  for the signal and noise contributions of the detector response, Eqs.~(\ref{eq:current-signal}) and (\ref{eq:current-noise}) respectively, as well as the Caves lower bound on the quantum noise (\ref{eq:current-min}). We consider Duffing nonlinearities $K_d=-3.4\times 10^{-6}$  and  $K_d=0$ (i.e., no nonlinearity). The integrated signal and noise bandwidth is taken to be $\delta\omega=2R_{\gamma}\gamma_{bm}$.  The corresponding parameter values are: probe line impedance $Z_p=50~{\mathrm{Ohms}}$, transmission line mode angular frequency $\omega_T/(2\pi)=5\times 10^9~{\mathrm{s}}^{-1}$, transmission line mode quality factor $Q_T=\omega_T/(2\gamma_{pT})=300$, mechanical frequency $\omega_m/(2\pi)=4\times 10^6~{\mathrm{s}}^{-1}$, mechanical quality factor $Q_m=\omega_m /(2\gamma_{bm})=10^3$, oscillator mass $m=10^{-16}~{\mathrm{kg}}$, Josephson junction critical current $I_{c}=4.5\times10^{-6}~\mathrm{A}$,  junction capacitance $C_{J}=10^{-14}~\mathrm{F}$, external flux bias $\Phi_{\mathrm{ext}}=0.442\ \Phi_0$, and external field in the vicinity of the mechanical resonator $B_{\mathrm{ext}}=0.05~\mathrm{T}$. These values give a zero-point uncertainty $\Delta_{zp}=1.45 \times 10^{-13}~{\mathrm{m}}$ and transmission line-oscillator coupling $K_{Tm}=1.1\times10^{-5}$.  

\begin{figure}[htbp]
\begin{center}
\includegraphics[width=4.0in]{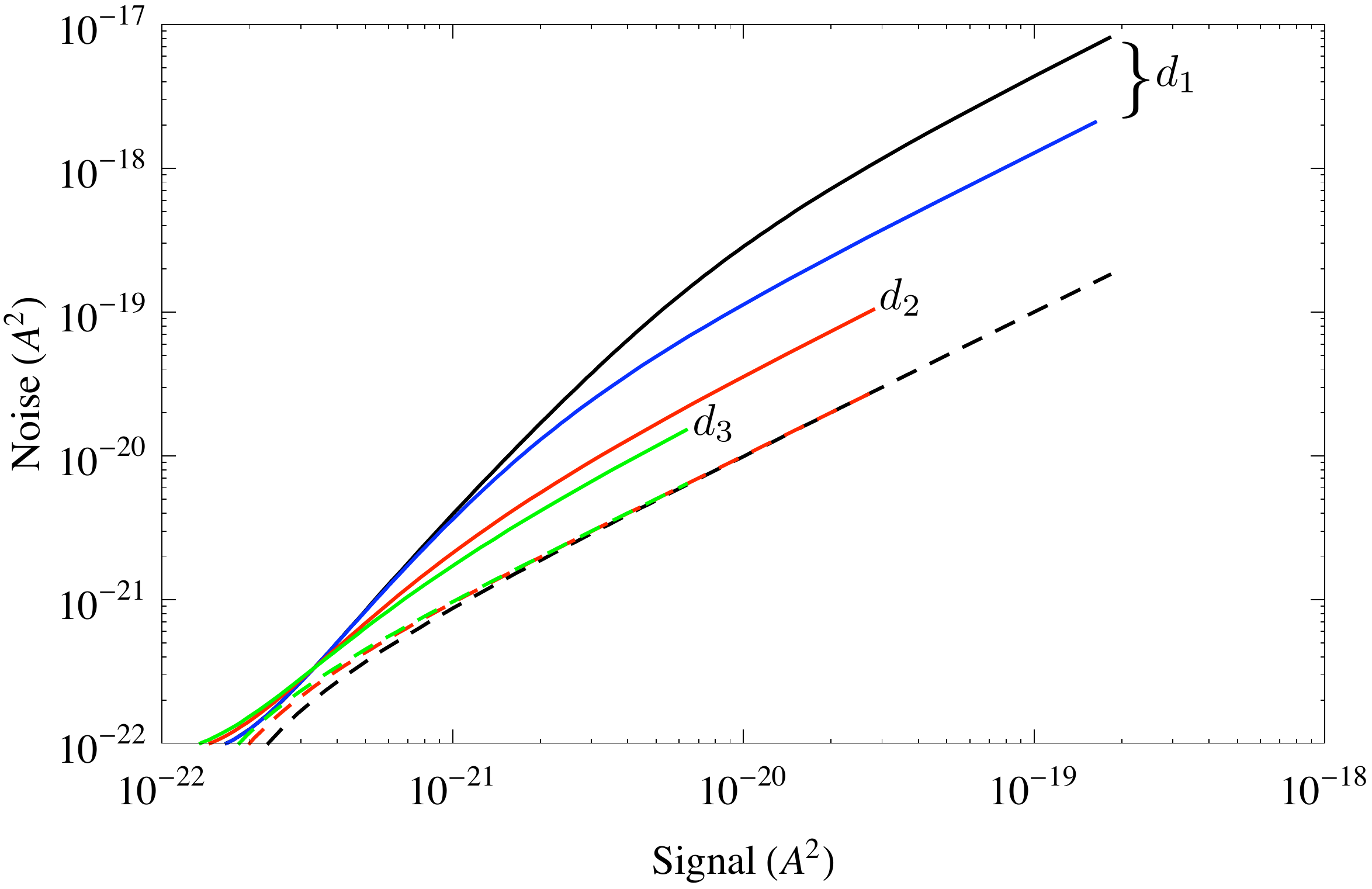}
\caption{Detector noise versus signal response at $\Delta\omega=0$ for harmonic ($K_d=0$) transmission line, Duffing ($K_d<0$) transmission line ($d_1$), and Caves' bound (black-dashed).  Noise for the nonlinear transmission line is also evaluated for blue detunings: $\Delta\omega=+0.2|\Delta\omega_{bi}|$ ($d_2$), and $+0.4|\Delta\omega_{bi}|$ ($d_3$). The labeled curves correspond to the  traces in Fig.~\ref{fig:region}.  The dashed, colored lines  give Caves' bound for the corresponding detuning values.}
\label{fig:detection}
\end{center}
\end{figure}

The advantage of using a spring softening nonlinearity, $K_d<0$, is clearly evident in Fig. \ref{fig:detection}, where we plot the noise versus response signal under increasing current drive for a transmission line both with and without Duffing term driven on resonance, $\Delta\omega=\omega_p-\omega_T=0$. We also plot the response of the nonlinear transmission line for several positively detuned values, $\Delta\omega=\omega_p-\omega_T>0$.  Termination of the curves indicates the signal value at which the damping renormalization $R_{\gamma}= 0$, beyond which the derived solutions become unphysical due to the net mechanical damping rate becoming negative and hence the motion unstable about the original fixed point.
Note that the same criterion, namely $R_{\gamma}>0$, is employed throughout the paper in order to ensure stability of the system. Again, the semiclassical, mean field approximation is expected to break down in the vicinity of termination points, where large fluctuations in the mechanical oscillator amplitude occur.  In Fig.~\ref{fig:detection}, we see that with positive detuning,  we can further approach the Caves bound. However, this is at the expense of reduced gain; depending on one's point of view, large renormalizations of the mechanical oscillator damping rate (and frequency) due to detector back action may or may not be allowed in detector displacement sensitivity figures of merit, affecting the maximum achievable gain as one approaches more closely the Caves bound.

\begin{figure}[htbp]
\begin{center}
\includegraphics[width=5.5in]{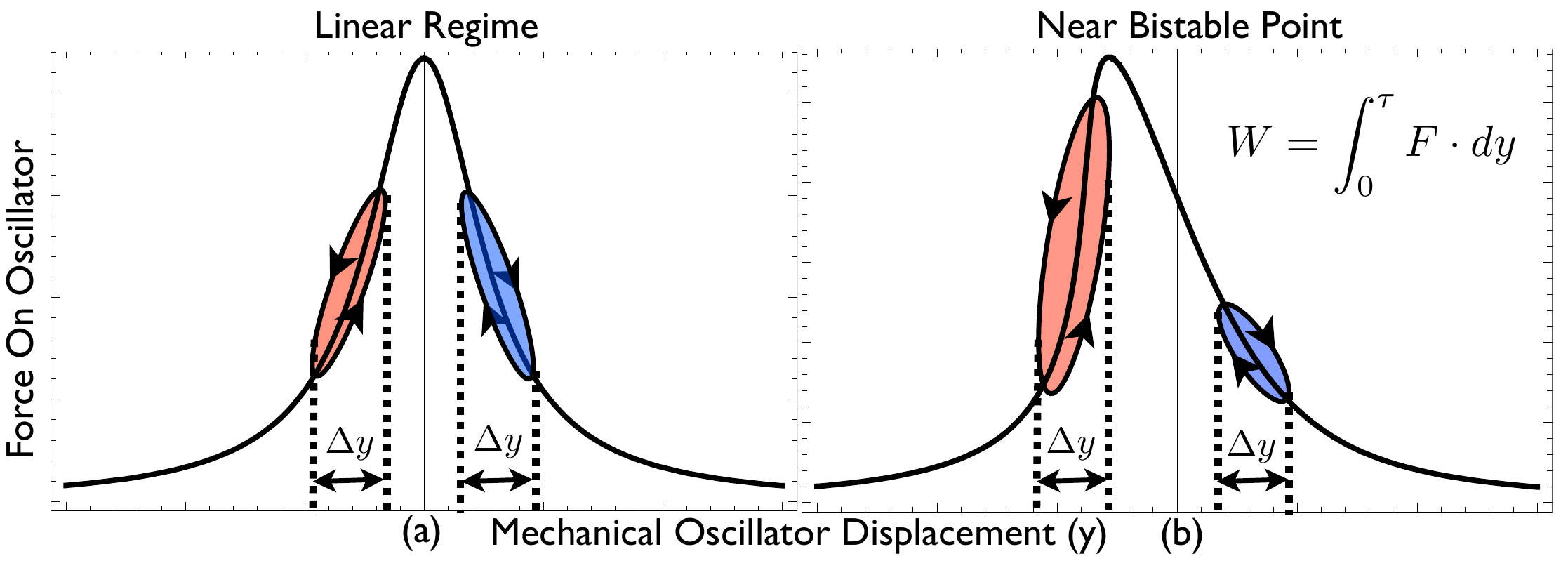}
\caption{Cartoon indicating the `radiation pressure' force exerted on the mechanical oscillator by the transmission line mode during one cycle of mechanical motion: (a) the harmonic transmission line mode approximation;  (b) approaching the onset of bistability.  The work done on the oscillator is proportional to the area swept out during each cycle, considerably exaggerated here for clarity.  Positive mechanical damping (red detuning) results on the positive slope side of the curves. Negative mechanical damping (blue detuning) results on the negative slope side. A spring softening nonlinearity can result in improved cooling for red detuning and improved signal-to-noise amplification  for blue detuning. }
\label{fig:curves}
\end{center}
\end{figure}

The trends displayed in Fig.~\ref{fig:detection} can be partly explained by invoking Fig.~\ref{fig:curves}, which indicates qualitatively the force on the mechanical oscillator due to the microwave transmission line `ponderomotive radiation pressure' force, both with vanishing  and with nonzero Duffing nonlinearity  and also for `red' and `blue'  pump frequency detunings.  
The work done on the mechanical oscillator by the radiation pressure force during one period of motion, due to the delayed transmission line resonator response, is given by the area enclosed within the hysteresis loop\cite{Kippenberg:2007p1995,Marquardt:2008p1309} and can be related to the steady-state back action damping rate through
\begin{equation}
\gamma_{\mathrm{back}}=-\frac{W}{\bar{E}}\frac{1}{\tau},
\label{eq:hysteresis}
\end{equation}
where $W$ is the work done on the mechanical oscillator, $\bar{E}$ is the average oscillator energy and $\tau$ is the period of motion.
When frequency pulling is taken into account,  the usual notions of red-detuned ($\Delta\omega<0$) or blue-detuned ($\Delta\omega>0$)  hold only in the weak drive limit.  We will assume red (blue)-detuned  to correspond to drive and detuning values $\Delta\omega$ where the net work done on the oscillator is negative (positive) as seen in Fig.~\ref{fig:curves}.
For a harmonic transmission line and for low drive powers, the frequency pulling effects can be ignored, since the effective Duffing coupling Eq.~(\ref{eq:k}) is proportional to the square of the transmission line-mechanical oscillator coupling $K_{Tm}$, which contributes only weakly for the considered parameter values.  Conversely, the Duffing term causes frequency pulling even at low input power and can significantly alter the slope of the response curve.  From Eq.~(\ref{eq:hysteresis}), the decreased slope on the blue detuned side leads to a decrease in the  damping rate magnitude which, through Eq.~(\ref{eq:noisetosignal}), leads as demonstrated above to a closer approach to the Caves' limit. As mentioned above, benefits in lower noise-to-signal resulting from further detuning deep into the blue region are offset by diminished achievable signal gain levels.   

Tuning the sign of the Duffing coupling $\mathcal{K}$ (\ref{eq:k}) to be positive, so that we have a hardening spring, results in an increased back action damping rate for blue detuning, and hence  a corresponding decrease in signal to noise relative to the harmonic transmission line resonator detector case.   

\section{\label{sec:cooling} Cooling}

Referring to the parametrizations (\ref{eq:paramcurrentsignal}) and (\ref{eq:paramcurrentnoise}) of the  signal and noise components of the detector response,  we define the mechanical oscillator's net occupation number through the following equation:
\begin{equation}
\gamma_{\mathrm{net}}(2n^{\pm}_{\mathrm{net}}+1)=\gamma_{bm}\left[2n(R_{\omega}\omega_{m})+1\right]+\gamma_{\mathrm{back}}(2n^{\pm}_{\mathrm{back}}+1),
\label{eq:gammanetnnet}
\end{equation}
 where the net damping rate is $\gamma_{\mathrm{net}}=\gamma_{bm}+\gamma_{\mathrm{back}}=R_{\gamma}\gamma_{bm}$. 
 The oscillator's net occupation number is then
 \begin{equation}
 2n_{\mathrm{net}}+1=R_{\gamma}^{-1}\left[2n(R_{\omega}\omega_{m})+1\right] +(1-R_{\gamma}^{-1}) (2n^{\pm}_{\mathrm{back}}+1).
 \label{eq:nnet}
 \end{equation}  
In order to  cool a mechanical oscillator to its ground state using detector back action, we therefore require a large detector back action damping rate, equivalently large damping rate renormalization $R_{\gamma}\gg 1$, together with a small detector back action effective occupation number $n^{\pm}_{\mathrm{back}}\ll 1$.   
 
\begin{figure}[htbp]
\begin{center}
\includegraphics[width=4.0in]{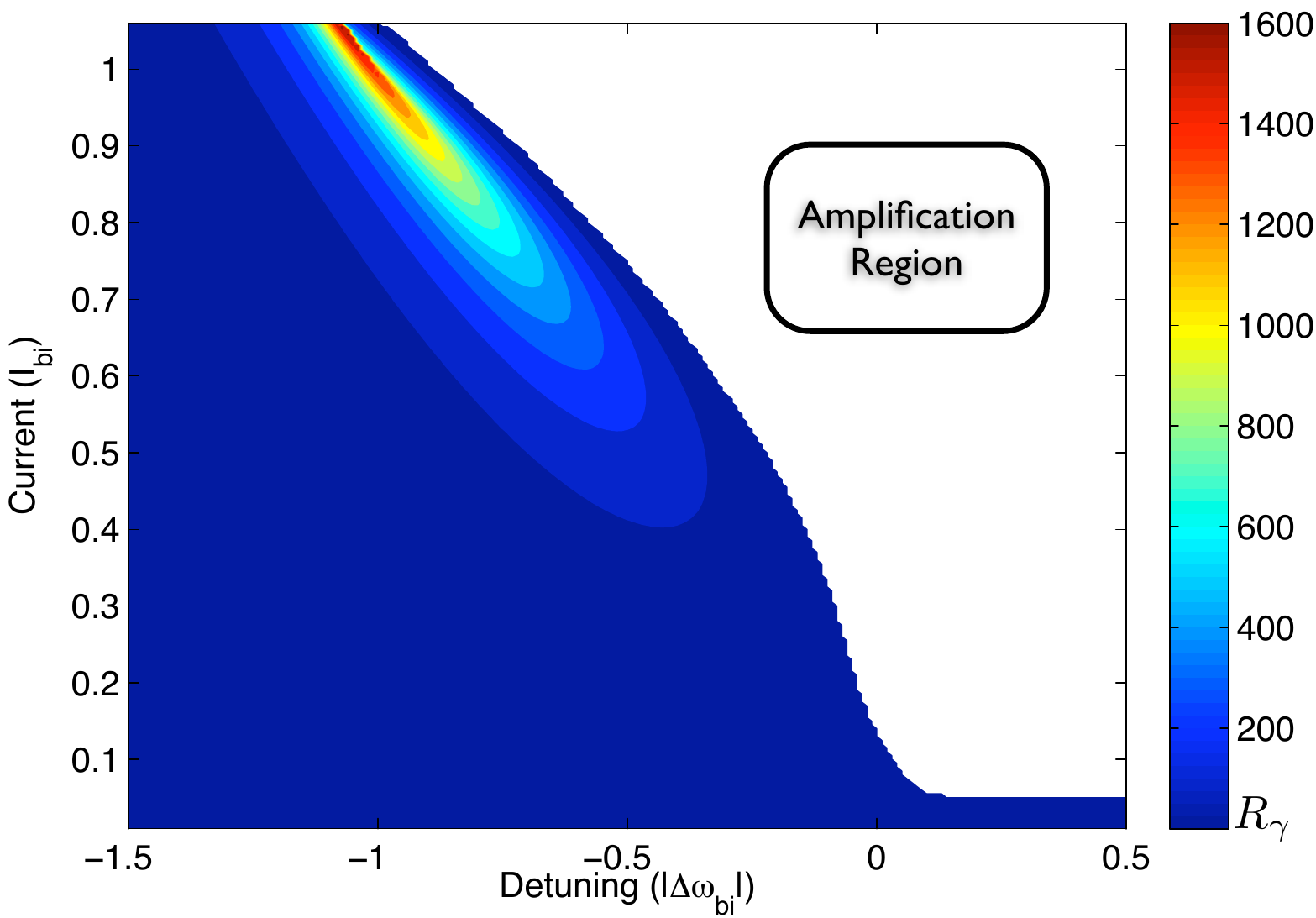}
\caption{Mechanical oscillator damping renormalization factor $R_{\gamma}$ for detunings both above and below the bistable detuning $\Delta\omega_{bi}$. The amplification region corresponds to negative back action damping, i.e., $R_{\gamma}<1$.}
\label{fig:gamma}
\end{center}
\end{figure}

Referring to Fig.~\ref{fig:curves}, operating closer to the bistability increases the negative work done per cycle on the oscillator by the cavity  and hence  increases the back action damping rate for given current drive.  In Fig.~\ref{fig:gamma}, we plot the mechanical oscillator damping rate renormalization factor $R_{\gamma}$, using the same parameter values as in  Sec.~\ref{sec:detection} (e.g., Duffing coupling $K_{d}=-3.4\times10^{-6}$), but with a larger yet still feasible mechanical quality factor $Q_m=10^4$ (which we shall adopt throughout this section).  We clearly see the enhanced damping as one approaches the onset of bistability given by $I_{{bi}}$ (\ref{eq:Ibi}) and $\Delta\omega_{{bi}}<0$ (\ref{eq:detune}).  

For the  example parameter choices of  Sec.~\ref{sec:detection}, we have $\omega_{m}/\gamma_{pT}\approx0.5$ and thus we are operating in the so-called bad cavity limit, where cooling close to the ground state (i.e., $n_{\mathrm{net}}\ll 1$) is not possible.\cite{Blencowe:2007p285,Marquardt:2007p978,WilsonRae:2007p502}
 While it is not difficult to achieve the good cavity limit $\omega_m>\gamma_{pT}$ simply by realizing sufficiently large quality factor superconducting microwave resonators, together with high frequency mechanical resonators,\cite{Teufel:2008p2398} it is nevertheless worthwhile to address how nonlinearities can improve on the cooling limits in the bad-cavity case. With the fundamental motivation to demonstrate macroscopic quantum behavior, the anticipated trend is to work with increasingly massive and hence lower frequency oscillators, making it progressively more difficult to achieve the good cavity limit.   

\begin{figure}[htbp]
\begin{center}
\includegraphics[width=6.0in]{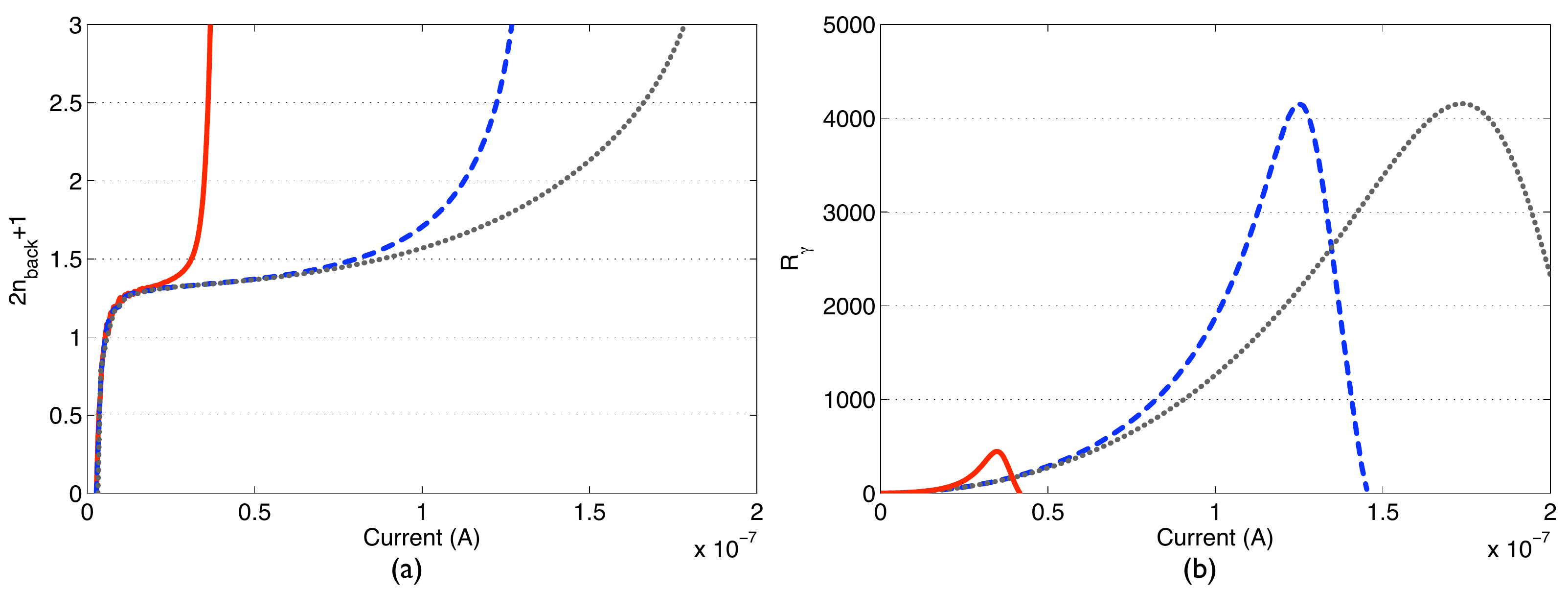}
\caption{(a) Detector noise effective back action occupation number versus current drive when red-detuned at $\Delta\omega=-\sqrt{\omega_{m}^{2}+\gamma_{pT}^{2}}$, $|\Delta\omega|<|\Delta\omega_{bi}|$, with a Duffing nonlinearity (solid line),  without  a Duffing nonlinearity (dashed line), and without both Duffing and ponderomotive nonlinearities (dotted line). (b) Oscillator coupling renormalization factor $R_{\gamma}$ for the corresponding back-action occupation number curves. These plots are obtained for the straight line trace labeled $c_1$ in Fig.~\ref{fig:region}.}
\label{fig:optimal}
\end{center}
\end{figure}

In Fig.~\ref{fig:optimal}, we plot the dependence of detector's noise effective back action occupation number $n_{\mathrm{back}}$ on microwave drive current amplitude  at the detuning bias $\Delta\omega=-\sqrt{\omega_{m}^{2}+\gamma_{pT}^{2}}$, where  $|\Delta\omega|<|\Delta\omega_{bi}|$. This is the optimum detuning in the harmonic, transmission line oscillator approximation, i.e.,  when nonlinear effects are ignored.    The noise effective occupation number is indicated  for both a nonzero ($K_d=-3.4\times10^{-6}$) as well as zero ($K_d=0$)  Duffing nonlinearity transmission line. We also show for comparison the effective back action occupation number when the frequency pulling effects of both the ponderomotive coupling $K_{Tm}$ and Duffing coupling  $K_{d}$ are neglected.  The latter case is obtained by dropping the nonlinear microwave mode amplitude term in the mean field equation~(\ref{eq:mean-field0}).  The sharp rise in occupation number and associated sharp drop in damping renormalization at larger current drives is a consequence of  crossing over into the amplification region due to negative frequency pulling of the cavity response relative to the fixed detuning.  The decrease in  occupation number as $I \rightarrow 0$ is accompanied by weak back action damping,  which prevents cooling the mechanical oscillator to  such occupation numbers. Note that at smaller current drives the damping renormalization in the presence of a Duffing nonlinearity peaks above the corresponding damping renormalization without the Duffing nonlinearity. This damping enhancement can be qualitatively explained with the aid of Fig.~\ref{fig:curves}. In the presence of the nonlinearity then, improved cooling can be achieved for smaller current drives.        

\begin{figure}[htbp]
\begin{center}
\includegraphics[width=3.2in]{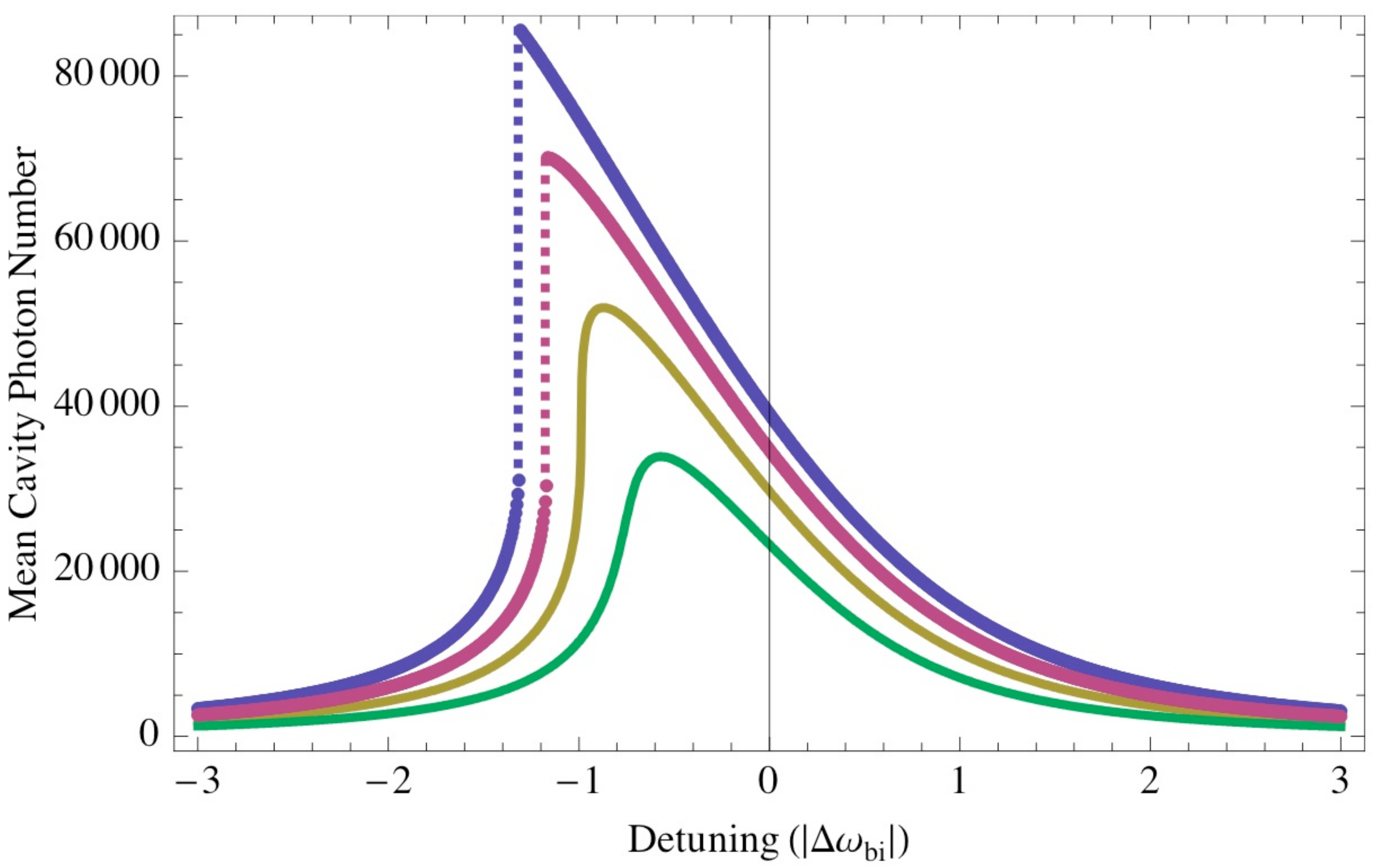}
\caption{Transmission line resonator response curve for $Q_{T}=300$ restricted to the small amplitude solution branch.  The example drive currents are $I/I_{bi}=0.8$ (green), 0.95 (yellow), 1.15 (red), and 1.3 (blue).  The  jump between small (red-detuned) and large (blue-detuned) amplitude solutions is indicated by the dotted lines.}
\label{fig:response}
\end{center}
\end{figure}

According to the above discussion, any  improvements in mechanical oscillator cooling are due solely to enhancements in the detector's back action damping rate for given drive; as can be seen from Fig.~\ref{fig:optimal}, the absolute minimum attainable detector effective occupation number is the same both in the presence and absence of the transmission line resonator Duffing nonlinearity.  While the effects of enhanced back action damping may be beneficial in situations where one is facing constraints on the maximum achievable drive power,\cite{Teufel:2008p2398} it would nevertheless be more significant if reductions in detector effective occupation number  could similarly  be achieved through nonlinear effects.   To see how this might be possible, we consider detunings corresponding to the pump frequency being to the left and away from the cavity resonance, i.e., $|\Delta\omega| > |\Delta\omega_{bi}|$, $\Delta\omega<0$. For such detunings, the mechanical oscillator `sees' a transmission line resonator effective quality factor that is determined by the steeper slope on the left side of the  response curve  (see Fig.~\ref{fig:curves}).  As we drive the transmission line resonator towards the lower bistable boundary (see Fig.~\ref{fig:region}), the slope of the response curve increases sharply and mimics a resonator with larger quality factor, effectively getting closer to the good cavity limit and hence resulting in a lower detector occupation number.\cite{Blencowe:2007p285,Marquardt:2007p978,WilsonRae:2007p502} Continuing to drive the transmission line resonator into the bistable region, and assuming that the resonator can be maintained on the low amplitude, red-detuned solution branch,\cite{Naaman:2008} the detector effective occupation number further decreases while the back action damping rate on the mechanical resonator increases (as explained by Fig.~\ref{fig:curves}). Eventually, the transmission line resonator becomes unstable at  the upper bistable boundary indicated in Fig.~\ref{fig:region}, and the oscillator jumps to the larger amplitude, blue-detuned solution (see Fig.~\ref{fig:response}).

\begin{figure}[htbp]
\begin{center}
\includegraphics[width=6.0in]{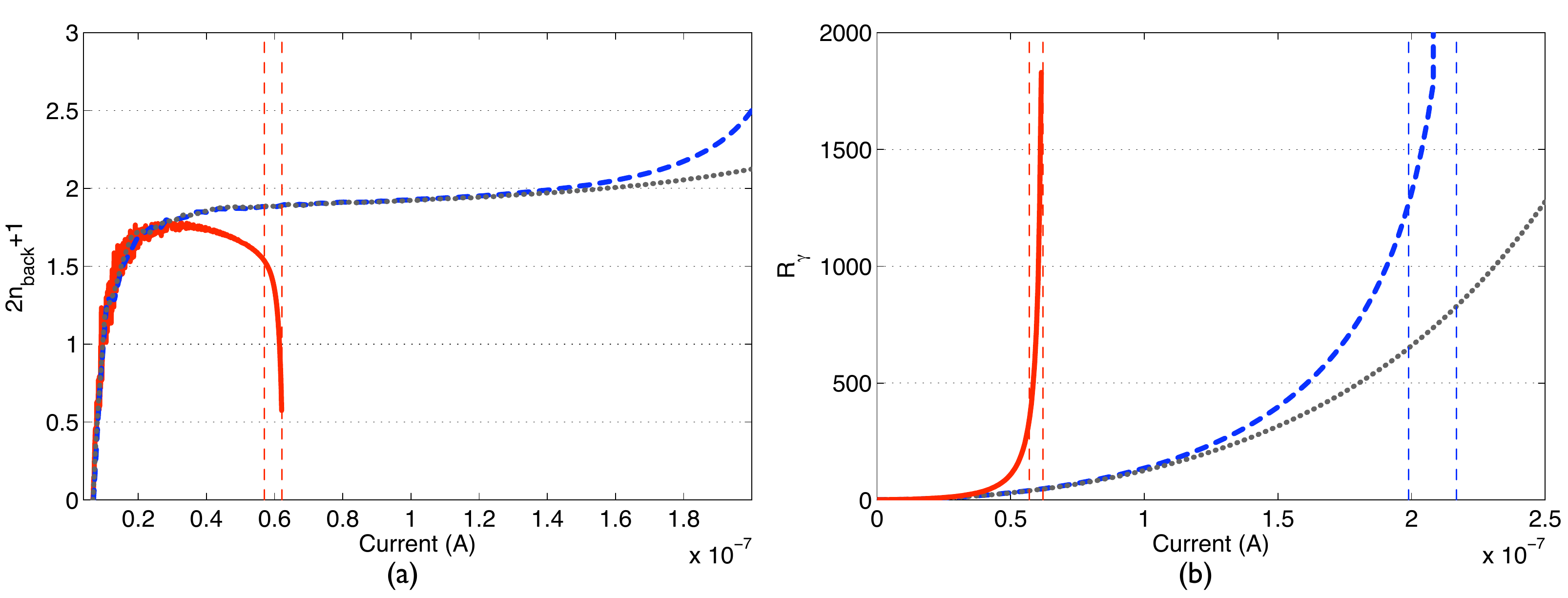}
\caption{(a) Detector noise effective occupation number versus current drive when red detuned at $\Delta\omega=1.3\ \Delta\omega_{bi}$, corresponding to straight line trace $c'_1$ in Fig.~\ref{fig:region}.  The Duffing nonlinear transmission line resonator occupation number (solid line) rapidly decreases as the resonator is driven towards the upper bistable boundary, assuming the resonator can be maintained on the small amplitude metastable stable solution branch. In contrast, a harmonic transmission line resonator (dashed line) or a cavity with neither Duffing nor ponderomotive nonlinearities in its mean field microwave mode equations (dotted line)  shows no such decrease in the occupation number. (b) Mechanical oscillator damping renormalization factor for the  same fixed detuning and drive current range.  The dashed vertical lines indicate the boundaries of the bistable region for the given transmission line resonator parameters.}
\label{fig:goodcavity}
\end{center}
\end{figure}

\begin{figure}[htbp]
\begin{center}
\includegraphics[width=6.0in]{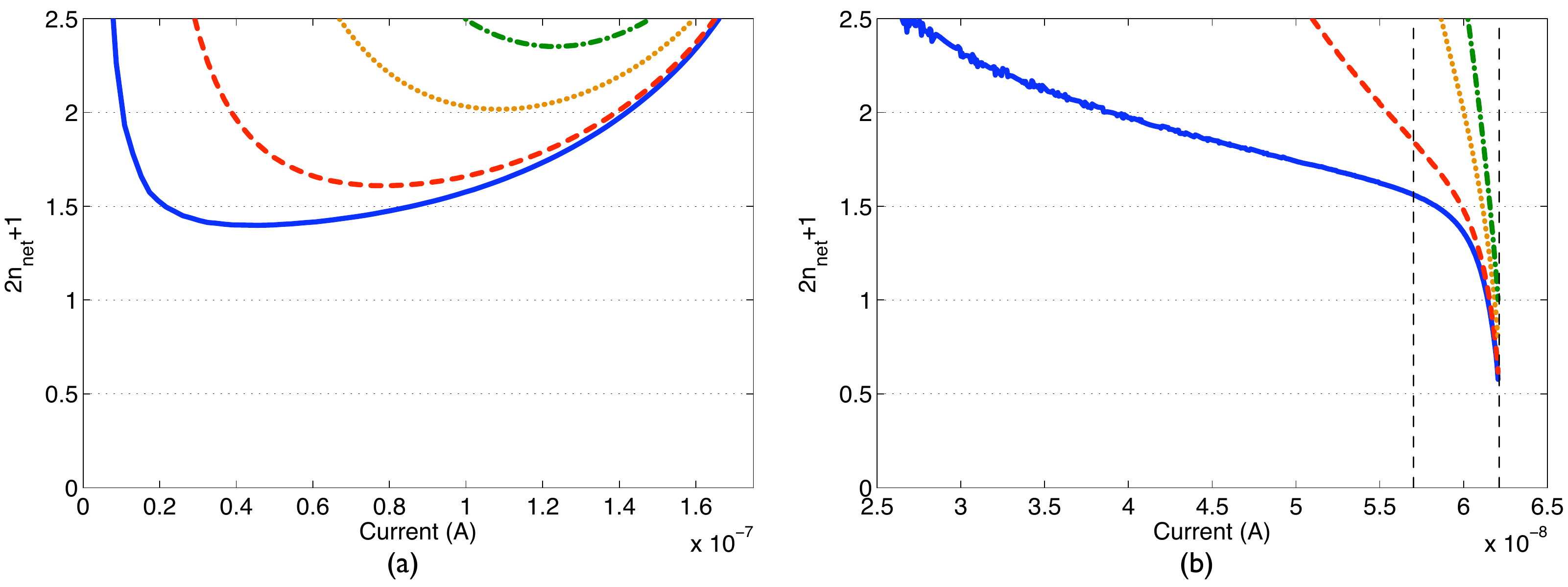}
\caption{(a) Net mechanical occupation number at $\Delta\omega=-\sqrt{\omega_{m}^{2}+\gamma_{pT}^{2}}$ for a harmonic transmission line resonator.  External bath temperatures: $T=1$ (solid line), 10 (dashed line), 50 (dotted line) and 100 (dot-dashed line) $\mathrm{mK}$. (b) Dependence of the net mechanical oscillator occupation number on current drive for a Duffing transmission line with detuning $\Delta\omega=1.3\ \Delta\omega_{bi}$.  The bistable region boundaries are indicated by the dashed vertical lines.}
\label{fig:cooling}
\end{center}
\end{figure}

In Fig.~\ref{fig:goodcavity}, we plot the dependence of the detector effective occupation number $n_{\mathrm{back}}$ on current drive for an example detuning value of $\Delta\omega=-2\sqrt{\omega_{m}^{2}+\gamma_{pT}^{2}}=1.3\Delta\omega_{bi}$.  
Driving the nonlinear transmission line resonator towards the upper boundary of the bistable region (see Fig.~\ref{fig:curves}) produces a sharp decrease in detector occupation number, and an occupation number value of $(2n_{\mathrm{back}}+1)\approx 0.55$ can be obtained,  well below that achievable when ignoring frequency pulling effects.  The harmonic cavity shows no such decrease in occupation number, indicating the qualitatively different quantum dynamical dependencies on $K_{d}$ and $K_{Tm}$ and the necessity of the former.  We can quantify the effect of frequency pulling by comparing with a harmonic transmission line resonator with a quality factor  $Q_{T}^{\mathrm{eff}}$ value chosen so as  to give the same detector effective occupation number.  For  the occupation number value $(2n_{\mathrm{back}}+1)\approx 0.55$, we have $Q_{T}^{\mathrm{eff}}\approx 600$,  corresponding to $\omega_{m}/\gamma^{\mathrm{eff}}_{pT}=0.95$, and therefore the mechanical oscillator behaves as if it is coupled to a cavity with double the quality factor.  This translates into lower net mechanical temperatures as shown in Fig.~\ref{fig:cooling}, where we give the net oscillator occupation number $n_{\mathrm{net}}$ (\ref{eq:nnet}) for various external bath temperatures.
The combination of nonlinearly-enhanced coupling $R_{\gamma}\gamma_{bm}$ and enhanced transmission line effective quality factor can be seen to significantly affect cooling of the mechanical motion, even for relatively large external temperatures.

In the numerical solutions to Eqs.~(\ref{eq:current-signal}) and (\ref{eq:current-noise}), the Lorentzian parametrizations (\ref{eq:paramcurrentsignal}) and (\ref{eq:paramcurrentnoise}) were found to give good approximations even when the upper bistable boundary is  approached quite closely. This is a consequence of the wide separation in the relaxation rates that determine the line widths of the harmonic transmission line resonator and unrenormalized mechanical oscillator modes, i.e., $\gamma_{bm}\ll\gamma_{pT}$. The upper bistable boundary has to be approached pretty closely in order for the nonlinear transmission line resonator ring-down time to exceed the renormalized mechanical oscillator damping time,  resulting in the breakdown of the  effective thermal description of the detector back reaction. In all of the plots shown in this section, the Lorentzian approximation is a good one over the resolvable scale of the plots. The actual minimum temperature  that can be achieved depends on the upper drive threshold where the Lorentzian approximation breaks down, as well as on the ability to keep the transmission line resonator on the small amplitude solution branch; the latter condition becomes progressively more difficult to satisfy as the upper boundary is approached, owing to the increasing probability of noise-induced jumps to the large amplitude branch.

\begin{figure}[htbp]
\begin{center}
\includegraphics[width=6.0in]{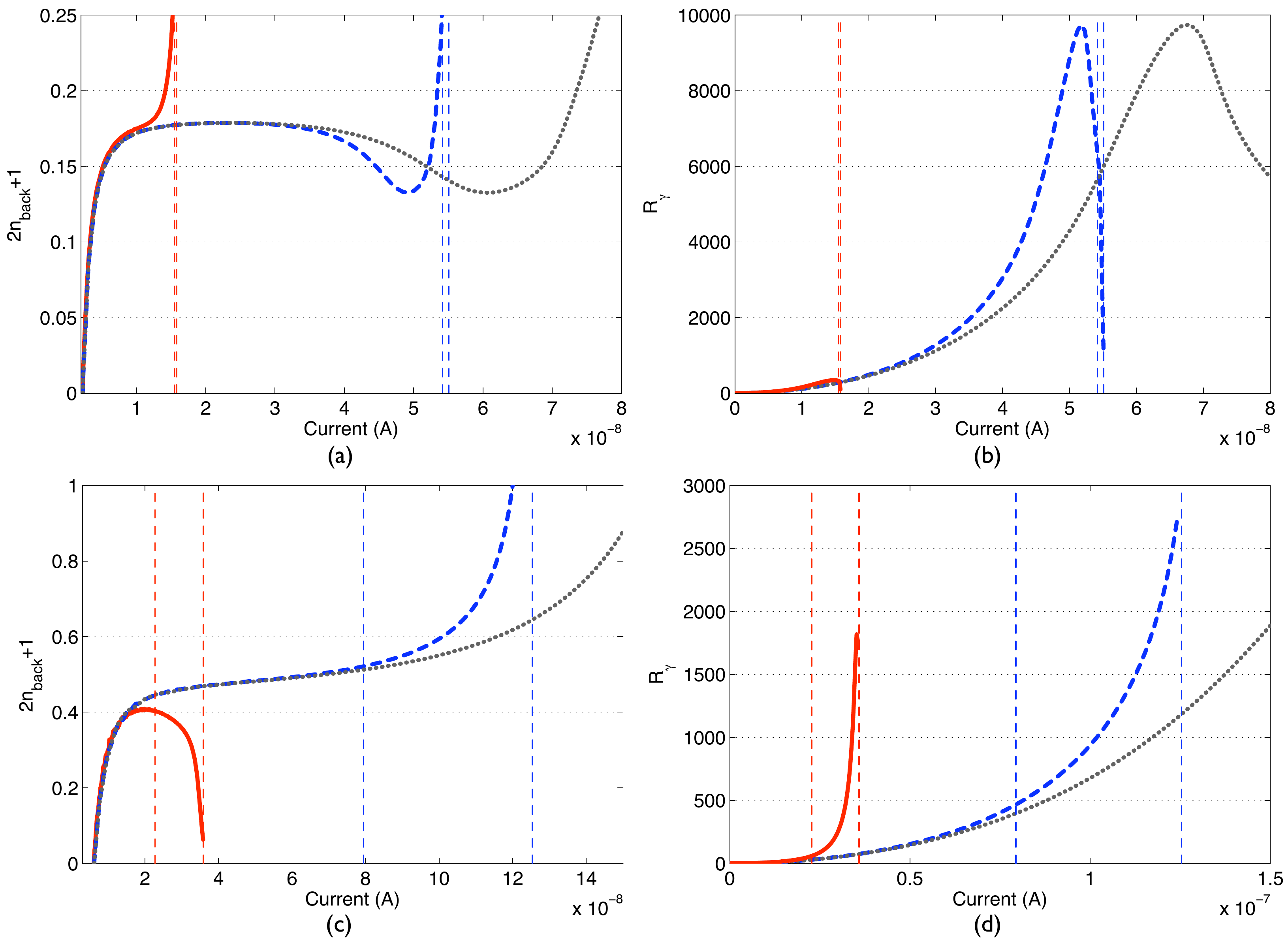}
\caption{(a) Detector noise effective occupation number versus current drive when red-detuned at $\Delta\omega=-\sqrt{\omega_{m}^{2}+\gamma_{pT}^{2}}$, $|\Delta\omega|<|\Delta\omega_{bi}|$, for a Duffing nonlinear (solid line),  harmonic (dashed line) transmission line, and with the effects of frequency pulling due to both ponderomotive and Duffing nonlinearities neglected (dotted line).  The vertical dashed lines give the bistable region boundaries for the Duffing and harmonic transmission line resonators. This plot is obtained for the straight line trace labeled $c_2$ in Fig.~\ref{fig:region} (b) Oscillator coupling renormalization factor $R_{\gamma}$ for optimal harmonic detuning.  (c) Detector occupation number detuned at twice the harmonic optimum, $\Delta\omega=2.2\Delta\omega_{bi}$ and locked to the lower stable amplitude solution. This plot is obtained for the straight line trace labeled $c'_2$ in Fig.~\ref{fig:region} (d) Corresponding back-action damping rate when driven to the upper bistable boundary.}
\label{fig:1000}
\end{center}
\end{figure}

\begin{figure}[htbp]
\begin{center}
\includegraphics[width=6.0in]{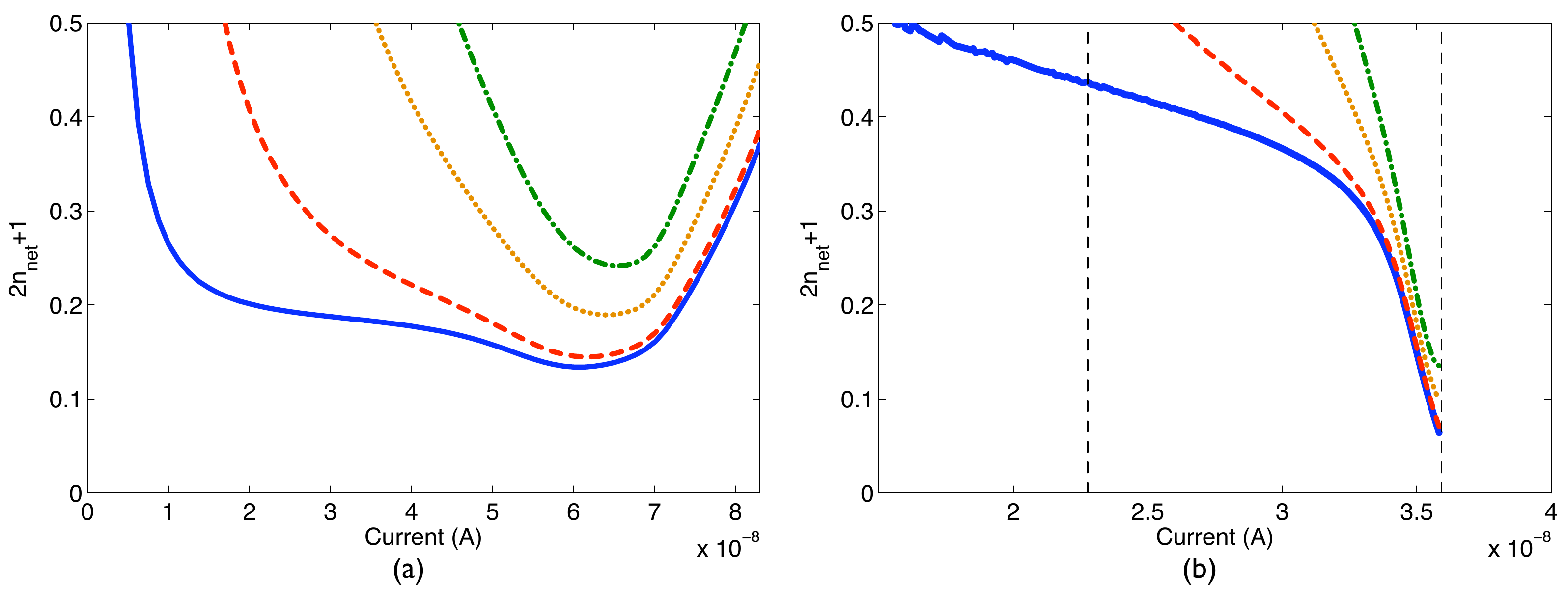}
\caption{(a) Net mechanical occupation number at $\Delta\omega=-\sqrt{\omega_{m}^{2}+\gamma_{pT}^{2}}$ for a harmonic transmission line resonator.  External bath temperatures: $T=1$ (solid line), 10 (dashed line), 50 (dotted line) and 100(dot-dashed line) $\mathrm{mK}$. (b) Dependence of the net mechanical oscillator occupation number on current drive for a Duffing transmission line with detuning $\Delta\omega=2.2\Delta\omega_{bi}$.}
\label{fig:cooling1000}
\end{center}
\end{figure}

A Duffing transmission line resonator nonlinearity can also produce cooling gains in the good cavity limit.  In Fig.~\ref{fig:1000}, we consider a transmission line resonator with $Q_{T}=1000$, giving $\omega_{m}/\gamma_{pT}=1.6$, and compare the  nonlinear transmission line resonator with the harmonic resonator approximation at optimal harmonic detuning.  Again, by detuning to twice the optimal harmonic resonator  value, $\Delta\omega=-2\sqrt{\omega_{m}^{2}+\gamma_{pT}^{2}}\approx2.2 \Delta\omega_{bi}$, we see that the effective back action occupation number decreases, while the  back action damping increases as the system is driven towards the upper boundary of the bistable region. 
Driving a Duffing transmission line resonator at twice the optimal harmonic detuning can yield a detector occupation number $(2n_{\mathrm{back}}+1)\approx 0.06$ just below the upper boundary of the bistable region, which is equivalent to an effective harmonic resonator quality factor of $Q_{T}^{\mathrm{eff}}\approx 1400$ or $\omega_{m}/\gamma^{\mathrm{eff}}_{pT}=2.2$.  In comparison, the minimum effective detector occupation number ignoring nonlinear effects is $2n_{\mathrm{back}}+1=0.13$.  In Fig.~\ref{fig:cooling1000},  we plot the net mechanical occupation number for the good cavity transmission line resonator both in the presence and absence of the Duffing nonlinearity.
Again, we see the strong cooling effects provided by frequency pulling of the cavity response. As discussed above, the minimum achievable net occupation number will depend on the threshold drive for which the Lorentzian approximation breaks down, as well as on the ability to lock the transmission line resonator onto the small amplitude solution branch in the bistable region.

\section{\label{sec:conclusion}Conclusions}

We have provided a quantum analysis of a nonlinear microwave amplifier for displacement detection and cooling of a mechanical oscillator. The amplifier comprises a microwave stripline resonator with embedded dc SQUID.  The SQUID gives rise to an effective, Duffing-type nonlinearity in the fundamental microwave mode equations, as well as a ponderomotive-type coupling between the microwave and fundamental mechanical modes.  It was found that  a spring-softening Duffing nonlinearity enables  a closer approach to the standard quantum limit for position detection as expressed by the Caves bound, as well as cooling closer to the  mechanical mode ground state. These findings can be qualitatively explained by considering the effects of frequency pulling in the response curve of the transmission line resonator `ponderomotive force' acting on the mechanical oscillator   (see Fig.~\ref{fig:curves}).  With blue detuning, the decrease in  damping allows for a closer approach to the quantum limit with large amplifier gain.  Conversely,  red detuning towards the bistable point of the force response curve increases the back action damping, improving the thermal contact to the detector `cold load'. Furthermore, effectively increasing the cavity quality factor due to the nonlinearity mimics the so-called good cavity limit in the harmonic case,  allowing cooling closer to the ground state. 

The present investigation has by no means exhaustively searched the large parameter space of  the transmission line resonator-embedded SQUID-mechanical resonator system for establishing the optimal displacement detection sensitivity and cooling parameters. Rather, our intention has been to point out general trends, using specific parameter values as illustrative examples. It may be that other choices of parameters (e.g., using a mechanical resonator with a smaller quality factor) lead to a closer approach to the standard quantum limit, or to cooling closer to the ground state.  

The  semiclassical, mean field methods  employed in the present work do not take into account  classical or quantum noise-induced jumps between the small and large amplitude metastable solutions that  become more likely as the bistability region boundaries are approached. Unless ways can be found to keep the transmission line resonator locked onto the smaller amplitude solution branch,\cite{Naaman:2008} the predicted effects of nonlinearity-induced cooling will be less substantial, as it will be necessary to operate deeper in the bistability region to avoid jumps. The driven microwave mode amplitude dynamics in the vicinity of the bistable region boundaries is still a relatively unexplored area that requires more sophisticated theoretical techniques in order to elucidate the fluctuations  between the small and large amplitude metastable solution branches.\cite{Dykman:1980p480,Dykman:2004p061102,Dykman:2005p021102,Dykman:2007p1864,Serban:2007p3199,Lifshitz:2007p040404,Kogan:2008p0972} This will be the subject of a future investigation.

\section*{Acknowledgements}
We thank A. Armour, M. Dykman, and R. Lifshitz for helpful discussions. This work was partly supported by the US-Israel Binational Science Foundation (BSF), the National Science Foundation (NSF) under NIRT grant CMS-0404031 (M.P.B.), and by the Israeli Science foundation (ISF), the Deborah
Foundation and the Ministry of Science (E.B.).

\appendix
\section{\label{sec:appendix}}
In this appendix, we give the derivation of the signal $a_T^{(1)}(\omega)$ and noise  $a_T^{(0)}(\omega)$ terms. Suppressing the signal dependent term $A(\omega,\omega')$ in  Eq.~(\ref{eq:atwonly}), we obtain the noise equation
\begin{eqnarray}\label{eq:zeroth-atw}
	{a}_{T}^{(0)}(\omega)&=&\int_{-\infty}^{\infty}d\omega'B(\omega,\omega'){a}_{T}^{(0)}(\omega-\omega')\cr
	&\times&\int_{-\infty}^{\infty}d\omega''\left[{a}_{T}^{(0)}(\omega''){a}_{T}^{{(0)}+}(\omega''-\omega')+{a}_{T}^{{(0)}+}(\omega''){a}_{T}^{(0)}(\omega''+\omega')\right]\cr
&+&D(\omega)\int_{-\infty}^{\infty}d\omega''\int_{-\infty}^{\infty}d\omega'
{a}_{T}^{{(0)}+}(\omega''){a}_{T}^{(0)}(\omega'){a}_{T}^{(0)}(\omega+\omega''-\omega')+C(\omega).
\end{eqnarray}
Keeping only terms to first order in $A(\omega,\omega')$ or $a_T^{(1)}$ in Eq.~(\ref{eq:atwonly}), we obtain the signal equation
 \begin{eqnarray}\label{eq:first-atw}
{a}_{T}^{(1)}(\omega)&=&\int_{-\infty}^{\infty}d\omega'{a}_{T}^{(0)}(\omega-\omega')A(\omega,\omega')+\int_{-\infty}^{\infty}d\omega'B(\omega,\omega'){a}_{T}^{(1)}(\omega-\omega')\cr
&\times&\int_{-\infty}^{\infty}d\omega''\left[{a}_{T}^{(0)}(\omega''){a}_{T}^{{(0)}+}(\omega''-\omega')+{a}_{T}^{{(0)}+}(\omega''){a}_{T}^{(0)}(\omega''+\omega')\right]\cr
&+&\int_{-\infty}^{\infty}d\omega'B(\omega,\omega'){a}_{T}^{(0)}(\omega-\omega')\int_{-\infty}^{\infty}d\omega''\left[{a}_{T}^{(1)}(\omega''){a}_{T}^{{(0)}+}(\omega''-\omega')\right.\cr
&+&\left.{a}_{T}^{{(1)}+}(\omega''){a}_{T}^{(0)}(\omega''+\omega')
+{a}_{T}^{(0)}(\omega''){a}_{T}^{{(1)}+}(\omega''-\omega')+{a}_{T}^{{(0)}+}(\omega''){a}_{T}^{(1)}(\omega''+\omega')\right]\cr
&+&D(\omega)\int_{-\infty}^{\infty}d\omega''\int_{-\infty}^{\infty}d\omega'\left[{a}_{T}^{{(0)}+}(\omega''){a}_{T}^{(0)}(\omega'){a}_{T}^{(1)}(\omega+\omega''-\omega')\right.\cr
&+&\left.{a}_{T}^{{(0)}+}(\omega''){a}_{T}^{(1)}(\omega'){a}_{T}^{(0)}(\omega+\omega''-\omega')+{a}_{T}^{{(1)}+}(\omega''){a}_{T}^{(0)}(\omega'){a}_{T}^{(0)}(\omega+\omega''-\omega')\right].
\end{eqnarray}
Decomposing $a_T^{(0)}(\omega)=\langle a_T^{(0)}(\omega)\rangle +\delta a_T^{(0)}(\omega)$ and expanding Eq.~(\ref{eq:zeroth-atw}) to first order in the quantum noise fluctuation $\delta a_T^{(0)}(\omega)$, we obtain the following two equations:
\begin{eqnarray}\label{eq:zeroth-atw-coherent}
\langle{a}_{T}^{(0)}(\omega)\rangle&=&\int_{-\infty}^{\infty}d\omega'B(\omega,\omega')\langle{a}_{T}^{(0)}(\omega-\omega')\rangle\cr
&\times&\int_{-\infty}^{\infty}d\omega''\left[\langle{a}_{T}^{(0)}(\omega'')\rangle\langle{a}_{T}^{{(0)}+}(\omega''-\omega')\rangle+\langle{a}_{T}^{{(0)}+}(\omega'')\rangle\langle{a}_{T}^{(0)}(\omega''+\omega')\rangle\right]\cr
&+&D(\omega)\int_{-\infty}^{\infty}d\omega''\int_{-\infty}^{\infty}d\omega'\langle{a}_{T}^{{(0)}+}(\omega'')\rangle\langle{a}_{T}^{(0)}(\omega')\rangle\langle{a}_{T}^{(0)}(\omega+\omega''-\omega')\rangle\cr
&+&\langle C (\omega)\rangle
\end{eqnarray}
and
 \begin{eqnarray}\label{eq:zeroth-atw-quantum}
	&&\delta{a}_{T}^{(0)}(\omega)=\int_{-\infty}^{\infty}d\omega'B(\omega,\omega')\delta{a}_{T}^{(0)}(\omega-\omega')\cr
&\times&\int_{-\infty}^{\infty}d\omega''\left[\langle{a}_{T}^{(0)}(\omega'')\rangle\langle{a}_{T}^{{(0)}+}(\omega''-\omega')\rangle+\langle{a}_{T}^{{(0)}+}(\omega'')\rangle\langle{a}_{T}^{(0)}(\omega''+\omega')\rangle\right]\cr
&+&\int_{-\infty}^{\infty}d\omega'B(\omega,\omega')\langle{a}_{T}^{(0)}(\omega-\omega')\rangle
\int_{-\infty}^{\infty}d\omega''\left[\delta{a}_{T}^{(0)}(\omega'')\langle{a}_{T}^{{(0)}+}(\omega''-\omega')\rangle\right.\cr
&+&\left.\delta{a}_{T}^{{(0)}+}(\omega''-\omega')\langle{a}_{T}^{(0)}(\omega'')\rangle+\delta{a}_{T}^{{(0)}+}(\omega'')\langle{a}_{T}^{(0)}(\omega''+\omega')\rangle\right.\cr
&+&\left.\delta{a}_{T}^{(0)}(\omega''+\omega')\langle{a}_{T}^{{(0)}+}(\omega'')\rangle\right]\cr 
&+&D(\omega)\int_{-\infty}^{\infty}d\omega''\int_{-\infty}^{\infty}d\omega'\left[\delta{a}_{T}^{{(0)}+}(\omega'')\langle{a}_{T}^{(0)}(\omega')\rangle\langle{a}_{T}^{(0)}(\omega+\omega''-\omega')\rangle\right.\cr
&+&\left.\delta{a}_{T}^{(0)}(\omega')\langle{a}_{T}^{{(0)}+}(\omega'')\rangle\langle{a}_{T}^{(0)}(\omega+\omega''-\omega')\rangle+\delta{a}_{T}^{(0)}(\omega+\omega''-\omega')\langle{a}_{T}^{{(0)}+}(\omega'')\rangle\langle{a}_{T}^{(0)}(\omega')\rangle\right]\cr
&+&\delta C(\omega).
\end{eqnarray}

Assuming $\langle a_T^{(0)}(\omega)\rangle$ can be expressed approximately as a delta function, i.e., $\langle a_T^{(0)}(\omega)\rangle=\chi\delta(\omega-\omega_p)$, Eq.~(\ref{eq:zeroth-atw-coherent}) reduces to Eq.~(\ref{eq:mean-field0}) for $\chi$. The semiclassical approximation to Eq.~(\ref{eq:first-atw}) for $a_T^{(1)}(\omega)$, with  $a_T^{(0)}(\omega)$ replaced by $\langle a_T^{(0)}(\omega)\rangle=\chi\delta(\omega-\omega_p)$, then becomes 
\begin{eqnarray}\label{eq:atw_chi}
&&\left\{1-2\left|\chi\right|^{2}\left[B(\omega,0)+B(\omega,\omega-\omega_{p})+D(\omega)\right]\right\}{a}_{T}^{(1)}(\omega)\cr
&&-\chi^{2}\left[2B(\omega,\omega-\omega_{p})+D(\omega)\right]{a}_{T}^{{(1)}+}(2\omega_{p}-\omega)=\chi A(\omega,\omega-\omega_{p}).
\end{eqnarray}
In order to invert and obtain ${a}_{T}^{(1)}(\omega)$, we require a second, linearly independent equation that also depends on ${a}_{T}^{{(1)}+}(2\omega_{p}-\omega)$. Such an equation is obtained by making the replacement $\omega\rightarrow 2\omega_{p}-\omega$ in Eq.~(\ref{eq:atw_chi}) and then taking the adjoint:
\begin{eqnarray}\label{eq:atw_one_chi}
&&\left\{1+2\left|\chi\right|^{2}\left[B(\omega-2\Delta\omega,0)+B(\omega-2\Delta\omega,\omega-\omega_{p})+D(\omega-2\Delta\omega)\right]\right\}{a}_{T}^{{(1)}+}(2\omega_p-\omega)\cr
&&+\chi^{*2}\left[2B(\omega-2\Delta\omega,\omega-\omega_{p})+D(\omega-2\Delta\omega)\right]{a}_{T}^{{(1)}}(\omega)=-\chi^* A(\omega-2\Delta\omega,\omega-\omega_{p}).
\end{eqnarray}
Now inverting, we obtain
\begin{equation}
{a}_{T}^{(1)}(\omega)=\alpha_{1}(\omega)A(\omega,\omega-\omega_{p})+\alpha_{2}(\omega)A(\omega-2\Delta\omega,\omega-\omega_{p}),
\end{equation}
where
\begin{equation}
\alpha_{1}(\omega)=\mathcal{D}(\omega)^{-1}\left\{1+2\left|\chi\right|^{2}\left[B(\omega-2\Delta\omega,0)+B(\omega-2\Delta\omega,\omega-\omega_{p})+D(\omega-2\Delta\omega)\right]\right\}\chi,\label{eq:alpha1}
\end{equation}
\begin{equation}
\alpha_{2}(\omega)=-\mathcal{D}(\omega)^{-1}\left[2B(\omega,\omega-\omega_{p})+D(\omega)\right]\left|\chi\right|^{2}\chi\label{eq:alpha2}
\end{equation}
and the determinant is given by,
\begin{eqnarray}\label{eq:determinant}
\mathcal{D}(\omega)&=&\left\{1-2\left|\chi\right|^{2}\left[B(\omega,0)+B(\omega,\omega-\omega_{p})+D(\omega)\right]\right\}\cr
&\times&\left\{1+2\left|\chi\right|^{2}\left[B(\omega-2\Delta\omega,0)+B(\omega-2\Delta\omega,\omega-\omega_{p})+D(\omega-2\Delta\omega)\right]\right\}\cr
&+&\left|\chi\right|^{4}\left[2B(\omega,\omega-\omega_{p})+D(\omega)\right]
\left[2B(\omega-2\Delta\omega,\omega-\omega_{p})+D(\omega-2\Delta\omega)\right].
\end{eqnarray}
A similar approach is used to obtain $\delta{a}_{T}^{(0)}(\omega)$ from Eq.~(\ref{eq:zeroth-atw-quantum}),  giving
\begin{equation}
	\delta{a}_{T}^{(0)}(\omega)=\beta_{1}(\omega)\delta C(\omega)+\beta_{2}(\omega)\delta C^{+}(2\omega_{p}-\omega),
\end{equation}
with
\begin{equation}
\beta_{1}(\omega)=\mathcal{D}(\omega)^{-1}\left\{1+2\left[B(\omega-2\Delta\omega,0)+B(\omega-2\Delta\omega,\omega-\omega_{p})+D(\omega-2\Delta\omega)\right]\left|\chi\right|^{2}\right\}\label{eq:beta1}
\end{equation}
and
\begin{equation}
\beta_{2}(\omega)=\mathcal{D}(\omega)^{-1}\left[2B(\omega,\omega-\omega_{p})+D(\omega)\right]\chi^{2}.\label{eq:beta2}
\end{equation}
\bibliographystyle{apsrev}
\bibliography{squid_paperv10.bib}

\end{document}